# Dynamic locking of an interacting spin system via periodic driving

**Dionisio Cendoya[1,2,*], Lisandro Buljubasich[1,2,*,†], Eric G. Keeler[3], and Carlos A. Meriles[4,5,†]**

Periodic driving plays a central role in quantum control, but its application in interacting spin systems is often restricted to near-resonant conditions, where standard averaging techniques remain valid. Here we investigate how detuning from resonance can be used to dynamically spin-lock a dipolar-coupled ensemble. We show that the combination of offset and pulse structure generates an effective Rabi field with sharply structured amplitude and tilt. This behavior — supported by a semi-analytical framework, numerical simulations and experiment — enables new approaches to many-body system control, here exemplified via offset-induced reversible interconversion of Zeeman and dipolar order, and heterospin polarization transfer away from rf-field matching conditions. Further, we leverage artificial-intelligence-assisted sequence design to explore regimes with long control cycles — where average Hamiltonian theory breaks down, but effective locking persists — opening pathways to rich offset-dependent responses. These findings position offset-enabled dynamic locking as a promising tool for quantum sensing, energy transfer, and spin-order manipulation beyond traditional approaches.

Periodic driving has emerged as a powerful tool for controlling complex quantum systems, enabling the engineering of novel states of matter and the stabilization of otherwise fragile dynamics[1-5]. In many-body settings, time-periodic control underpins phenomena ranging from Floquet prethermalization[6-8] and discrete time crystals[9-11] to light-induced superconductivity[12], quantum transport[13-15], and dynamic localization[16]. These effects exploit the ability of periodic drives to reshape the effective Hamiltonian governing the system's evolution, sometimes allowing access to emergent behaviors with no equilibrium counterpart[17-19].

In a driven many-body system, the dynamics over one full period of duration $t_c$ are often captured stroboscopically by the unitary propagator $U(t_c) = \mathcal{T}\exp\left(-i\int_0^{t_c}\mathcal{H}(t)dt\right)$. In this expression, $\mathcal{H}(t)$ is a time-dependent Hamiltonian that includes the intrinsic interactions of the system as well as the periodic action of the external control pulses, and $\mathcal{T}$ denotes the time-ordering operator. In typical implementations, detuning between the drive and the system natural frequency is minimized, as increasing the offset leads to rapid degradation of controllability. Furthermore, the rotating-frame Hamiltonian becomes unbounded as the offset grows, which seriously complicates a perturbative description of the system's evolution, e.g., by recasting $U(t_c)$ through the Magnus expansion[7,20-23].

In this work, we investigate many-body spin dynamics under periodic driving in the presence of large, tunable offsets. Rather than treating detuning as a nuisance to be minimized, we introduce it deliberately as a control parameter and uncover a rich phenomenology that emerges beyond the regime where standard perturbative methods apply. Specifically, we study how interacting spin systems respond to periodic pulse sequences when the offset varies continuously, revealing a regime of dynamical spin locking in which spins synchronize to effective fields that are shaped jointly by the pulse structure and the detuning. We then leverage these dynamics to demonstrate robust manipulation of spin order, including new approaches to polarization transfer and adiabatic demagnetization in the rotating frame. Further, we resort to artificial-intelligence-optimized pulse sequences to explore longer-period dynamics, where the locking field acquires a sharply structured offset dependence with only mild detuning. These findings reveal an intriguing route to achieving long-lasting coherent control in interacting systems, with implications for quantum sensing, simulation, and the study of driven many-body dynamics.

To investigate the effects of detuning on periodically driven many-body systems, we study a prototypical ensemble of strongly interacting spins under coherent control. As a representative platform, we use a polycrystalline sample of adamantane — a solid-state system of dense dipolar-coupled protons[24] whose interactions quickly dephase any starting transverse magnetization (Fig. 1a). Our driving protocol consists of a repeated sequence of $\pi/2$-pulses about alternating axes, implemented with interpulse spacings $\tau$ and a stroboscopic readout every $6\tau$ (Fig. 1b). The dynamic spin locking cycle we use — which we refer to as DSL-4 — consists of four concatenated WaHuHa units[25] whose intended near-resonance effect is to average out the dipolar couplings while preserving frequency offsets. Formally, the evolution over a single cycle of duration $t_c$ is captured by the stroboscopic propagator $U(t_c)$, which, under standard average Hamiltonian theory, approximates to $U(t_c) \sim \exp(-i\,\mathcal{H}^{(0)} t_c)$; here, $\mathcal{H}^{(0)} = 2\pi\nu \cdot I_z/3$ is the effective Hamiltonian, $\nu$ denotes the offset relative to the rotating-frame Larmor frequency, and $I_z$ represents the global spin operator along the $z$-axis.

[1]Facultad de Matemática, Astronomía, Física y Computación, Universidad Nacional de Córdoba, Córdoba, Argentina. [2]CONICET, Instituto de Física Enrique Gaviola (IFEG), Córdoba, Argentina. [3]New York Structural Biology Center, New York, NY 10027, USA. [4]Department of Physics, CUNY- The City College of New York, New York, NY 10031, USA. [5]CUNY-Graduate Center, New York, NY 10016, USA. [*]Equally contributing authors. [†]E-mail: lisandro.buljubasich@unc.edu.ar, cmeriles@ccny.cuny.edu.

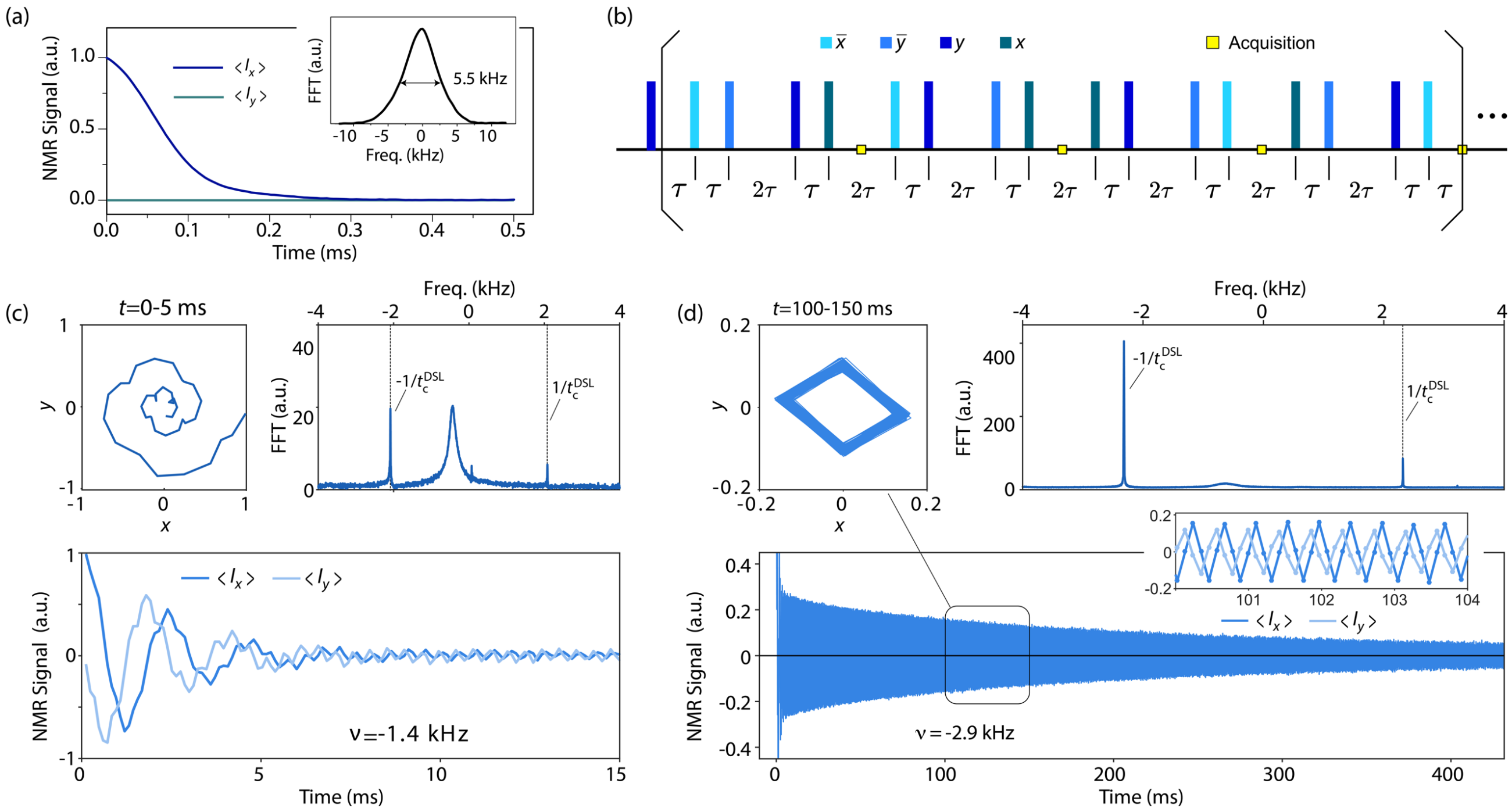

**Figure 1 | Dynamics of a many-body system under periodic driving.** (a) $^1$H nuclear magnetic resonance (NMR) signal from a sample of adamantane ($C_{10}H_{16}$) following a resonant $(\pi/2)_y$-pulse excitation; the insert displays the magnitude of the fast Fourier transform (FFT). (b) Schematics of the DSL-4 control protocol. All rf pulses (solid blocks) amount to $\pi/2$-rotations with relative phases denoted by colors; we implement stroboscopic sampling every $6\tau$. The unit in parenthesis represents a "cycle" of duration $t_c^{\mathrm{DSL}}$, which we recurrently repeat. (c) Measured evolution of the in-plane magnetization during application of the pulse train in (b) for a frequency offset $\nu = -1.4$ kHz. The left upper inset reproduces the measured trajectory of the global magnetization vector, $(\langle I_x\rangle, \langle I_y\rangle)$, throughout the observation window. The spectrum (upper right inset) shows a peak at the scaled frequency $\nu/3 = -0.47$ kHz. (d) Same as in (c) but for a frequency offset of -2.9 kHz. Following a short-lived transient, the magnetization decays slowly over a time window spanning a fraction of a second. The system dynamics locks into a nearly-closed orbit at a frequency $-1/t_c^{\mathrm{DSL}}$. In all experiments, the $\pi/2$-pulse duration is 3.6 µs, and the interpulse interval is $\tau = 20$ µs.

As shown in Fig. 1c for a near-resonant offset $\nu = -1.4$ kHz, the global magnetization exhibits oscillations at the expected scaled frequency $\nu/3$, as revealed by the spectrum in the inset. These oscillations reflect the linear scaling of frequency offsets under DSL-4, a key feature traditionally exploited for solid-state magnetic resonance spectroscopy[25,26]. Superimposed on this oscillatory response is a gradual decay, a consequence of residual dipolar couplings not fully averaged out by the pulse sequence (Supplementary Material (SM), Section I). This response, however, dramatically changes as the offset grows. Figure 1d shows an example at $\nu = -2.9$ kHz where the system locks into a long-lived, nearly-in-plane orbit with characteristic frequency set not by ν but by the cycle duration itself (Fig. 1d).

To understand the dynamics governing this many-body system, we first write the evolution operator over one DSL-4 cycle as[25]

$$U(t_c) = \prod_{k=1}^{n} P_k e^{-i\tau\mathcal{H}} = \prod_{k=1}^{n} e^{-i\tau\mathcal{H}_k'}. \qquad (1)$$

In the above expression, $n$ is the number of free-evolution time intervals $\tau$ within one cycle, $P_k$ represents the action of the radio-frequency (rf) pulse at the end of the $k$-th interval (or the identity if no pulse is present), and as usual, operators in a product must be assumed acting successively to the left in order of increasing index. The second equality recasts $U(t_c)$ in terms of the "toggling-frame" Hamiltonians $\mathcal{H}_k' = (\prod_{l=k}^{n} P_l)\mathcal{H}(\prod_{l=k}^{n} P_l)^{-1}$ using the fact that $\prod_{k=1}^{n} P_k = 1$ for DSL-4. For a system of protons connected via the dipolar coupling term $\mathcal{D}$, the rotating-frame spin Hamiltonian takes the form $\mathcal{H} =$ $\omega I_z + \mathcal{D}$, effectively unbounded if the offset $\omega = 2\pi\nu$ is allowed to grow arbitrarily large. Therefore, describing the right-hand side of Eq. (1) in terms of a Magnus expansion — the traditional treatment — becomes invalid unless $|\nu|$ is sufficiently small (less than ~1.5 kHz for the conditions in Fig. 1).

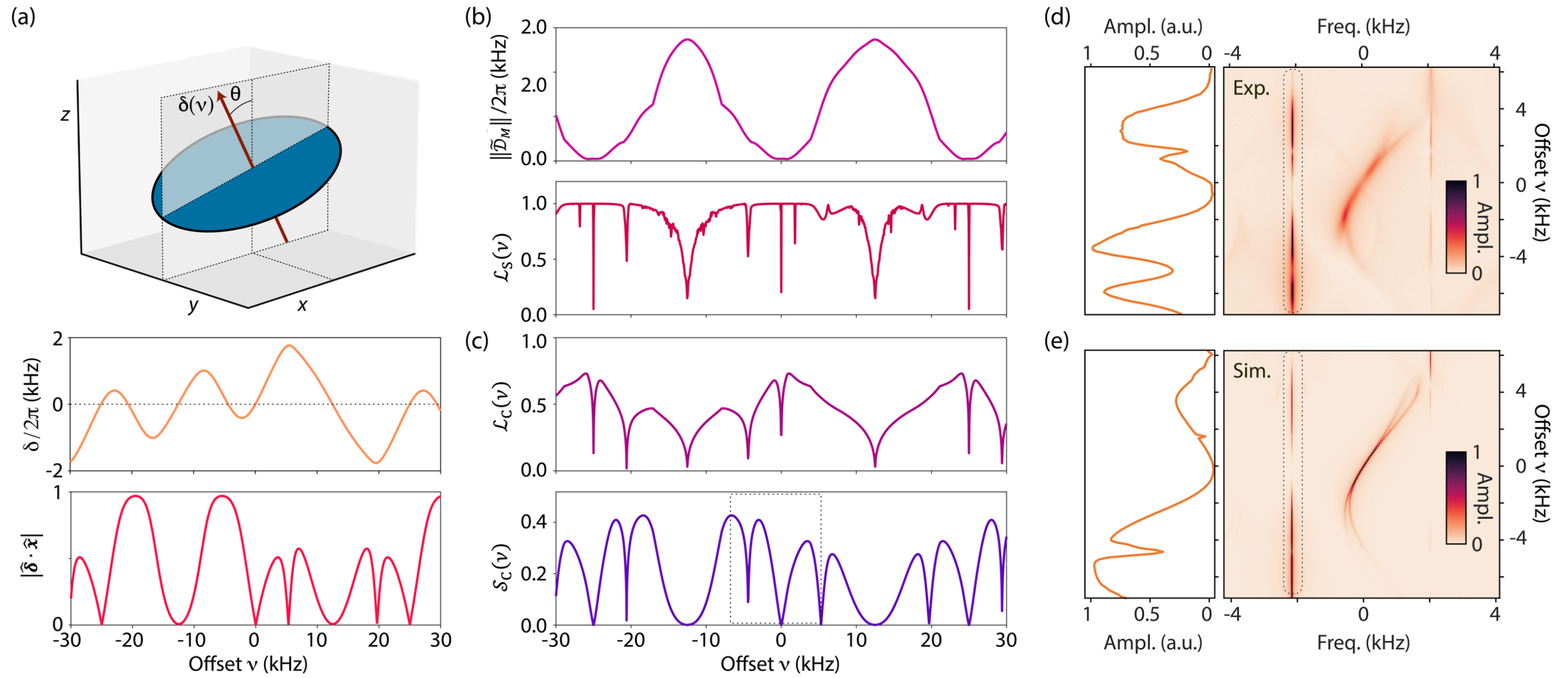

**Figure 2 | Offset-dependent structure of the locking field.** (a) (Top) Schematic representation of the effective locking field $\boldsymbol{\delta}$ in the rotating frame featuring offset dependent amplitude and projection angle (top and bottom plots, respectively). (b) Offset-dependent norm of the effective dipolar Hamiltonian $\|\widetilde{\mathcal{D}}_M\|$ over one DSL-4 cycle and simulated locking efficiency $\mathcal{L}_S(\nu)$ (top and bottom panels, respectively). (c) Calculated locking efficiency $\mathcal{L}_C(\nu)$ and signal amplitude $\mathcal{S}_C(\nu)$. We derive a highly structured response, with dips typically reflecting on unfavorable alignment or amplitude of the locking field. (d) Fourier transform of the measured DSL-4 NMR signal (absolute value mode) as a function of ν. The side plot shows the $^1$H resonance amplitude at $-1/t_c^{\mathrm{DSL}}$ as a function of the offset ν, qualitatively consistent with the predicted response (dashed square in the bottom panel of (c)). (e) Same as in (d) but derived from a simulation of a 10-spin cluster. The experimental conditions in (d) are those of Fig. 1.

It is possible, nonetheless, to gain insight on the system dynamics if one separates the action of the offset and dipolar terms; we get in this case

$$U(t_c) = \prod_{k=1}^{n} e^{-i\tau\widetilde{\mathcal{D}}_k'} \prod_{k=1}^{n} Q_k, \qquad (2)$$

where $\widetilde{\mathcal{D}}_k' = (\prod_{l=k}^{n} Q_l)\mathcal{D}_k'(\prod_{l=k}^{n} Q_l)^{-1}$ with $k = \{1 \ldots n\}$ are offset-dressed versions of the toggling-frame dipolar coupling operators, and $Q_k = \exp\left(-i\omega\tau I_{z,k}'\right)$ (see SM, Section I). Without loss of generality, the right-hand side product in Eq. (2) can be expressed as $\prod_{k=1}^{n} Q_k = \exp(-it_c\boldsymbol{\delta}(\omega)\cdot\mathbf{I})$, which can be seen as the evolution dictated by a locking field $\boldsymbol{\delta}$ featuring $\omega$-dependent Rabi amplitude and direction. Correspondingly, one expects a long-lasting, "spin-locked" signal for offset frequencies where $\boldsymbol{\delta}$ simultaneously displays strong amplitude and good alignment with the initial magnetization.

We note that the similarity with conventional spin-locking[27] — including pulsed spin locking[28-30] — is only nominal: Rather than aligning with a static rotating-frame direction, the magnetization locks into a closed orbit, as explicitly shown in Fig. 1d where the time separation between consecutive stroboscopic acquisitions is smaller than $t_c$. The emergence of dynamic locking — where the observable traces a large-amplitude, periodic trajectory synchronized with the pulse sequence — was noted in early studies of interacting spin systems in the perturbative regime[31] but, as we show next, its potential has remained largely untapped.

The key to this potential lies in the rich interplay between $\boldsymbol{\delta}$ and the offset-dressed dipolar interaction that surfaces as one considers detunings throughout the range defined by $\omega\tau = \pm\pi$, the point beyond which the behavior becomes cyclic. For DSL-4, $\boldsymbol{\delta}$ invariably lies within the $xz$-plane but its tilt and amplitude show a complex dependence with the offset (Fig. 2a). In particular, sufficiently large negative detuning causes $\boldsymbol{\delta}$ to align with $x$ — i.e., in a direction perpendicular to the effective field defined by $\mathcal{H}^{(0)}$ — which explains our ability to efficiently lock $I_x$ magnetization in Fig. 1d.

It follows from Eq. (2) that the sequence-induced dependence on the offset is also imprinted on the dipolar Hamiltonian, whose effective value can be extracted from the Magnus relation $\prod_{k=1}^{n}\exp\left(-i\tau\widetilde{\mathcal{D}}_k'\right) = \exp\left(-it_c\widetilde{\mathcal{D}}_M\right)$, applicable here at all detunings since $t_c\|\widetilde{\mathcal{D}}_k'\| \ll 1$. We capture the system dynamics in the bottom panel of

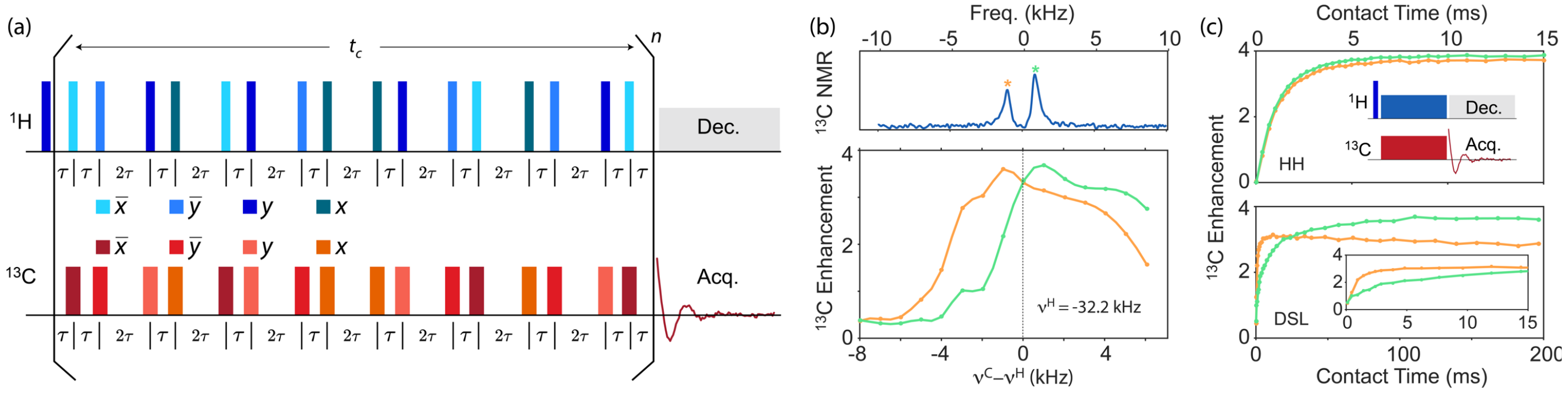

**Figure 3 | DSL-enabled heteronuclear polarization transfer.** (a) Simultaneous $^1$H and $^{13}$C dynamic spin locking. Both species are subjected to synchronized DSL-4 control with identical interpulse spacing and cycle structure, but with independent rf Rabi amplitudes ($\omega_1^H = 2\pi \times 143$ kHz, $\omega_1^C = 2\pi \times 85$ kHz), chosen far away from the Hartmann-Hahn (HH) condition. The acquisition (Acq.) is performed on the $^{13}$C channel following the transfer period during continuous $^1$H decoupling (Dec.). (b) Top: Representative $^{13}$C spectrum in adamantane featuring two $^{13}$C resonances from chemically inequivalent sites. Bottom: Extracted $^{13}$C polarization enhancement as a function of detuning difference; color-coded traces reflect the relative enhancement of either $^{13}$C resonance. (c) Bottom: Build-up of $^{13}$C polarization as a function of contact time $nt_c$ under DSL control with the $^{13}$C detuning chosen to optimize transfer to the higher-frequency $^{13}$C resonance; the inset shows the short-time dynamics (0–15 ms). Top: Reference $^{13}$C polarization build-up under standard HH conditions. During DSL, the $^1$H ($^{13}$C) $\pi/2$-pulse duration is 1.75 μs (2.95 μs) and the interpulse separation is $\tau = 4$ μs; the decoupling $^1$H Rabi field during acquisition is $\omega_1^H = 2\pi \times 143$ kHz. During HH, the Rabi amplitudes are $\omega_1^H = \omega_1^C = 2\pi \times 76$ kHz.

Fig. 2b, where we plot the locking efficiency $\mathcal{L}_S(\nu)$ as derived from a 10-spin cluster simulation where the magnetization initially aligns with the effective locking axis. This plot quantifies the competition between the effective locking field and the internal dipolar coupling that drives decoherence, and hence can be approximately described through a function $\mathcal{L}_C(\nu) = 1/(1 + r^\alpha)$; in this expression, we define $r(\nu) = \|\widetilde{\mathcal{D}}_M\|/|\delta|$, and $\alpha$ is a steepness parameter (top plot in Fig. 2c). Using $\hat{\boldsymbol{\delta}}$, $\hat{\boldsymbol{u}}$ to respectively denote the unit vectors along the locking field and initial magnetization, the dynamically locked signal amplitude can be expressed as $\mathcal{S}_C(\nu) = |\hat{\boldsymbol{\delta}} \cdot \hat{\boldsymbol{u}}|\, \mathcal{L}_C(\nu)$, plotted in the lower panel of Fig. 2c for initial magnetization along $x$. A detailed analysis shows that most dips in $\mathcal{L}(\nu)$ (and $\mathcal{S}(\nu)$) can be traced to the condition $|\boldsymbol{\delta} \cdot \hat{\boldsymbol{u}}| = m\pi$ with $m$ integer (SM, Section I). Systematic observations in adamantane as a function of offset as well as numerical modeling of a 10-spin cluster demonstrate good agreement with this model (Figs. 2d and 2e).

Having established a theoretical and experimental picture of dynamic spin locking, we now turn to illustrative applications enabled by this framework. Figure 3 presents one such example in the form of polarization transfer between two spin species. In contrast to the conventional Hartmann–Hahn (HH) mechanism — where the rf amplitudes scale inversely with the ratio between the spin gyromagnetic ratios — the present protocol operates far from this condition, with markedly different nutation (Rabi) frequencies. Indeed, the relevant quantity is the effective locking field, generated by the full control sequence and shaped by both detuning — the case in Fig. 3b — and interpulse spacing, independent of the gyromagnetic ratios. The distinct underlying mechanism manifests in the transfer dynamics (Fig. 3c): We observe a structured exchange process, where different resonances couple with different efficiencies. Such behavior suggests that the transfer pathway itself can be shaped, for instance, to favor specific spectral components through appropriate control design. Further, the approach decouples energy matching from rf power constraints, thus creating opportunities to render polarization transfer more versatile and less hardware-demanding. More broadly, this example highlights how periodically driven control can mediate energy and polarization flow in interacting systems. The same principles extend to other protocols, including adiabatic demagnetization in the rotating frame[27], extensively exploited in relaxometry applications[32] (SM, Section III).

To explore the generality of the locking phenomenon beyond DSL-4, we next examine the performance of AI-derived control schemes[33] (see also SM, Section IV). As an illustration, we focus on aDSL-67, an adaptive sequence characterized by longer pulse blocks, designed under the premise that less frequent repetition can yield solutions featuring a rich structure throughout a range of smaller, tunable offsets (Figs. 4a and 4b). In particular, the numerical example in Fig. 4c shows the magnetization settling into a quasi-circular orbit of near maximum amplitude for an offset as small as -733 Hz, revealed here via the sub-cycle sampling. To analyze its dependence with $\nu$, we again apply a frequency-domain approach (Figs. 4d and 4e). Unlike the case in Figs. 2d, 2e, we identify spin-locked peaks at multiples $q = \pm 1 \ldots \pm 4$ of the inverse cycle time, well within the frequency range where the response to detuning is nearly linear. We note that traditional control pulsed sequences are often too short to access these sharp resonances, making the locking behavior observed here an intriguing platform for $T_1$-limited magnetometry with dense ensembles of interacting spins[34,35].

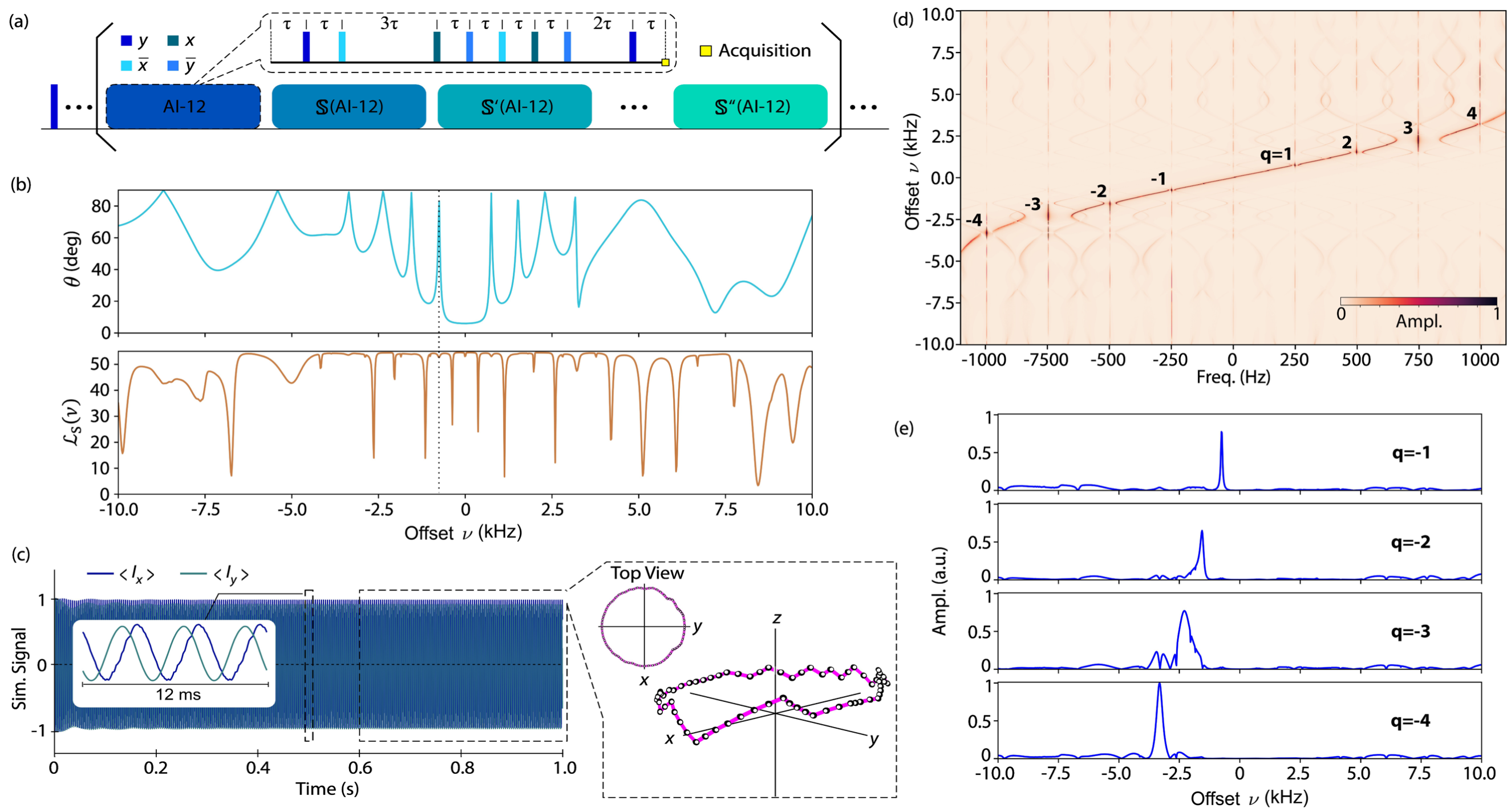

**Figure 4 | AI-assisted design of dynamic spin locking.** (a) AI-derived pulse protocol. Following excitation, we recurrently apply a 536-pulse block (square brackets) comprising 67 sub-units (rounded rectangles). These contains eight $\pi/2$-pulses (rectangles) arrayed in an AI-derived phase pattern, or a variant extracted from symmetry operations $\mathbb{S}$ (SM, Section IV); the total duration of the full block is $t_c^{\mathrm{AI}} = 804\tau$ and we set $\tau = 5$ µs. (b) Calculated locking field tilt angle and simulated locking efficiency $\mathcal{L}_S(\nu)$ (top and bottom plots, respectively) for the sequence in (a) using a 10-spin cluster. (c) Simulated spin cluster signal for $\nu = -733$ Hz. The inset is a zoomed-out view of the $\langle I_x \rangle$ and $\langle I_y \rangle$ components of the global magnetization. (Right inset) Trajectory of the magnetization vector $(\langle I_x \rangle, \langle I_y \rangle, \langle I_z \rangle)$ during the last 0.4 s of evolution (dashed square) corresponding to 100 aDSL-67 cycles; successive orbits fall exactly on each other, thus making only 67 points visible. (d) Fourier transform (absolute value mode) of the signal in (c) for variable offset $\nu$. Numbers on the 2D plot indicate multiples of the inverse block period. (e) Spectral amplitudes at multiples of the inverse period as derived from (d).

In sum, we have shown that a periodically driven spin system can exhibit long-lived, closed magnetization trajectories that remain recurrent even in the presence of strong dipolar interactions. These persistent orbits arise from a form of dynamic spin locking, a behavior in which the global observable traces a stable, periodic path defined by the pulse sequence. Our analytical treatment, based on toggling-frame Hamiltonians dressed by detuning, offers a predictive framework that captures the system dynamics across both conventional and AI-designed pulse protocols. Notably, the effective locking field $\delta(\omega)$ is independent of the spin's gyromagnetic ratio, allowing for species-agnostic control and cross-species polarization transfer far from conventional rf matching conditions. The versatility of this approach is underscored by the ability to generate resilient dynamics across different pulse structures, offering new avenues for coherent control in complex quantum systems with alternative interaction models, including those where suitable protocols are not yet known.

These features could be particularly impactful for quantum sensing, dynamic nuclear polarization, and multimodal spectroscopy, especially in resource-constrained or miniaturized settings. Specifically, the offset dependence of the stabilized dynamics opens new avenues for magnetometry, as the signal response with detuning can be made more pronounced through tailored control parameters. Harnessing such tunable features could transform interaction-driven correlations into a metrological resource. In the same vein, an open question is whether the resilience observed for global magnetization can be extended to more fragile, entangled states — and if so, whether tailored control sequences might enable their stabilization in a similar fashion.

## Data availability

The data that support the findings of this study are available from the corresponding author upon reasonable request.

## Code availability

All source codes for data analysis and numerical modeling used in this study are available from the corresponding authors upon reasonable request.

## Acknowledgments

We acknowledge helpful discussion with Fabián Vaca Chávez. We specially thank Keith Cannon for assistance in implementing the AI search. D.C. and L.B. acknowledge support from Secyt-UNC 32520230100029CB and CONICET PIP 11220200102451CO. C.A.M. acknowledges support from the U.S. National Science Foundation via grants NSF-2203904, NSF-2328993, and NSF-2506082; he also acknowledges access to the facilities and research infrastructure of the NSF CREST IDEALS, grant number NSF-2112550. C.A.M. is a member of the New York Structural Biology Center (NYSBC). Experimental data collection at NYSBC was supported by NIH grants S10OD034275 and P41GM118302.

## Author contributions

D.C. and L.B. implemented all simulations; they also carried out the experiments in collaboration with E.G.K.; D.C. carried out the AI-assisted search. D.C., L.B., and C.A.M. developed the analytical model. All authors analyzed the data; L.B and C.A.M. supervised the project. C.A.M. wrote the manuscript with input from all authors.

## Competing interests

The authors declare no competing interests.

## Correspondence

Correspondence and requests for materials should be addressed to L.B. and C.A.M.

Supplementary Material for

# Dynamic locking of an interacting spin system via periodic driving

**Dionisio Cendoya[1,2,*], Lisandro Buljubasich[1,2,*,†], Eric Keeler[3], and Carlos A. Meriles[4,5,†]**

*[1]Facultad de Matemática, Astronomía, Física y Computación, Universidad Nacional de Córdoba, Córdoba, Argentina.*
*[2]CONICET, Instituto de Física Enrique Gaviola (IFEG), Córdoba, Argentina.*
*[3]New York Structural Biology Center, New York, NY 10027, USA.*
*[4]Department of Physics, CUNY- The City College of New York, New York, NY 10031, USA.*
*[5]CUNY-Graduate Center, New York, NY 10016, USA.*

*[*]Equally contributing authors.*

[†]E-mail: lisandro.buljubasich@unc.edu.ar, cmeriles@ccny.cuny.edu.

**Table of content**

## I. Mathematical Framework

In this section, we develop a theoretical framework to describe the offset-dependent dynamics of a periodically driven dipolar spin ensemble. We first make use of standard average Hamiltonian theory (AHT), which provides an effective description under near-resonant conditions and weak dipolar coupling. This approach reveals how detuning modifies the leading-order average Hamiltonian, but we then show through numerical simulations that it fails to capture the long-lived coherent dynamics observed at larger offsets. To resolve this discrepancy, we introduce a modified treatment based on the separation of detuning and dipolar terms in the toggling frame, leading to a description in terms of a protocol-dependent locking field and offset-dressed dipolar couplings. This formalism, detailed below, captures the emergence of dynamic spin locking and provides an accurate, semi-analytical prediction of the system's response across a broad range of detunings.

When necessary, we complement the above description with numerical simulations of finite spin clusters with varying sizes and geometries. These models incorporate full pairwise dipolar couplings and allow us to probe how offset-dependent dynamics manifest in realistic many-body settings. Despite the necessarily low number of spins (typically 10 or 12), these simulations reveal nearly identical periodic responses across cluster types. Further, the overall agreement with experiment is good, confirming they provide a reasonably accurate platform to interpret our observations.

### I.1 Effective Hamiltonian from Average Hamiltonian Theory

To understand the role of detuning in our periodically driven many-body spin system, we begin by applying Average Hamiltonian Theory (AHT). This framework allows us to derive an effective time-independent Hamiltonian that governs the system's evolution under stroboscopic observation — i.e., at discrete times synchronized with the period of the external drive.

We consider a system of dipolarly coupled spin-½ particles with internal Hamiltonian

$$\mathcal{H} = \omega I_z + \sum_{i>j} 2\pi D_{ij}\left(I_z^{(i)} I_z^{(j)} - \frac{1}{4}\left(I_+^{(i)} \cdot I_-^{(j)} + I_-^{(i)} \cdot I_+^{(j)}\right)\right), \tag{1}$$

where $\omega = 2\pi\nu$ is the off-resonance frequency (controllable externally), $D_{ij}$ are the coupling constants, and we assume the applied magnetic field is sufficiently strong to truncate the dipolar Hamiltonian to its secular components. The operator $I_z = \sum_j I_z^{(j)}$ denotes the collective spin projection along the laboratory $z$-axis, with $I_z^{(j)}$ the Pauli-$z$ operator (in units of $\hbar = 1$) acting on spin $j$. Similarly, $I_+^{(j)}$ ($I_-^{(j)}$) is the raising (lowering) spin operator for the $j$-th particle. The system is subjected to a periodic radio-frequency (rf) drive in the form of pulse protocol with period $t_c$; in the particular case of DSL-4, $t_c = 24\tau$ where $\tau$ is the interpulse delay (Supplementary Fig. 1). The total Hamiltonian is

$$\mathcal{H}_{\mathrm{tot}}(t) = \mathcal{H} + \mathcal{H}_{\mathrm{rf}}(t), \tag{2}$$

where $\mathcal{H}_{\mathrm{rf}}(t)$ encodes the applied pulse sequence. Under periodic driving, the density matrix at a given time is defined by the evolution operator over one cycle, $U(t_c)$. Separating the action of the rf pulses $P_k$, we write

$$U(t_c) = \prod_{k=1}^{n} P_k e^{-i\tau\mathcal{H}} = \prod_{k=1}^{n} e^{-i\tau\mathcal{H}_k'} \prod_{k=1}^{n} P_k = \prod_{k=1}^{n} e^{-i\tau\mathcal{H}_k'}\ , \tag{3}$$

where $\mathcal{H}_k' = (\prod_{l=k}^{n} P_l)\mathcal{H}(\prod_{l=k}^{n} P_l)^{-1}$ is the toggling frame Hamiltonian during the $k$-th free evolution interval, and we have used $\prod_{k=1}^{n} P_k = 1$ for DSL-4. The traditional approach rewrites the right hand side of Eq. (3) in terms of an effective Hamiltonian, i.e.,

$$U(t_c) = \exp(-i\mathcal{H}_{\mathrm{M}} t_c). \tag{4}$$

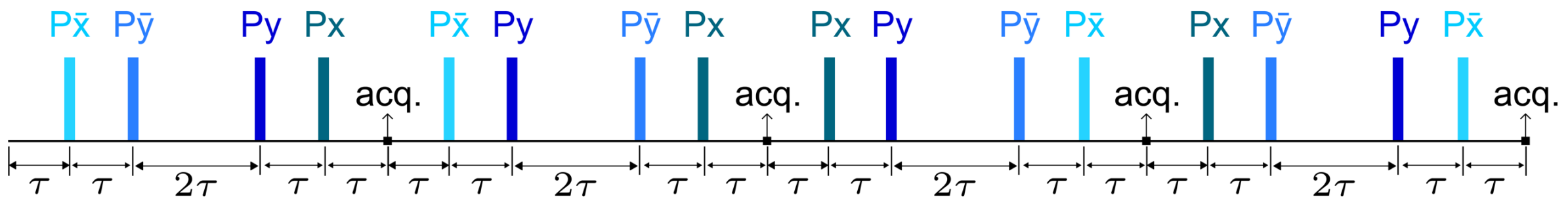

**Supplementary Figure 1 | DSL-4 pulse sequence.** Schematic representation of the pulse sequence.

In the above expression, $\mathcal{H}_{\mathrm{M}} = \sum_{l=0}^{\infty} \mathcal{H}^{(l)}$ takes the form of a series where each term can be obtained recursively through the Magnus expansion[1]. Typically, one can gain analytical traction by truncating the series to the first few orders — most notably the zeroth order $\mathcal{H}^{(0)}$ corresponding to the average Hamiltonian over the cycle — an approximation valid when the characteristic interactions are sufficiently small.

To assess the impact of frequency detuning in the spin dynamics we first consider the simpler case of an ensemble with negligible dipolar couplings (e.g., liquids, see below). From a detailed calculation that includes up to third order contributions, we write the Magnus Hamiltonian for the DSL-4 sequence as

$$\mathcal{H}_{\mathrm{M}}^{\mathrm{DSL}} \approx \mathcal{H}_{\mathrm{M}}^{(0)} + \mathcal{H}_{\mathrm{M}}^{(1)} + \mathcal{H}_{\mathrm{M}}^{(2)} + \mathcal{H}_{\mathrm{M}}^{(3)}, \tag{5}$$

where

$$\mathcal{H}_{\mathrm{M}}^{(0)} = \frac{1}{3}\omega I_z \ ; \qquad \mathcal{H}_{\mathrm{M}}^{(1)} = \tau\omega^2\left(-\frac{2}{3}I_x + \frac{1}{3}I_z\right) \ ;$$
$$\mathcal{H}_{\mathrm{M}}^{(2)} = \tau^2\omega^3\left(-\frac{4}{3}I_z\right) \ ; \qquad \mathcal{H}_{\mathrm{M}}^{(3)} = \tau^3\omega^4\left(\frac{19}{9}I_x + \frac{19}{18}I_z\right). \tag{6}$$

Together, Eqs. (5) and (6) imply the appearance of a detuning-dependent "locking field" whose amplitude — null on resonance — rapidly grows with the offset. Further, the orientation of this field — referred to as $\boldsymbol{\delta}(\omega)$ hereafter — slightly tilts away from the *z*-axis to gradually gain a small component along *x*. This phenomenon was already identified in early high-resolution NMR studies[2], where a small offset was observed to improve decoupling efficiency and to yield a long-lived, though weak, signal arising from the component of the initial magnetization aligned with $\boldsymbol{\delta}$. In this regime, the average Hamiltonian picture provides a useful intuitive description, but only for low detunings where higher-order terms remain small.

This simple picture soon proves insufficient, as shown in Supplementary Fig. 2 where we simulate the response of a non-interacting spin cluster under DSL-4 as we vary the frequency offset: We first characterize the exact dynamics in Supplementary Fig. 2a, where we plot the Fourier transform (magnitude mode) of the time-dependent response assuming initial magnetization along *y* and for detuning reaching up to ±3 kHz. We find that the linear frequency shift valid near resonance rapidly breaks down below about -500 Hz, at which point the dependence first becomes non-linear and subsequently inverts, steering towards zero-frequency with increasing detuning when $\nu \in [-1.5, -0.8]$ kHz. Comparison with the spectra obtained from successive Magnus expansions (Supplementary Figs. 2b through 2d) only agree near zero detuning and rapidly diverge as the offset grows. Even at third order, the effective Hamiltonian misses the rich spectral structure of the exact dynamics, underscoring that the locking-field intuition provided by AHT cannot be extrapolated beyond small offsets.

### I.2 Offset-Dressed Toggling Frame Description

To better capture the dynamics at play, we first note that

$$\mathcal{H}_k' = \left(\prod_{l=k}^{n} P_l\right)\mathcal{H}\left(\prod_{l=k}^{n} P_l\right)^{-1} = \omega I_{z,k}' + \mathcal{D}_k' \tag{7}$$

where we defined $I_{z,k}' = (\prod_{l=k}^{n} P_l) I_z (\prod_{l=k}^{n} P_l)^{-1}$, $\mathcal{D}_k' = (\prod_{l=k}^{n} P_l)\mathcal{D}(\prod_{l=k}^{n} P_l)^{-1}$, and $\mathcal{D} = \sum_{i>j} 2\pi D_{ij}\left(I_z^{(i)} I_z^{(j)} - \frac{1}{4}\left(I_+^{(i)} \cdot I_-^{(j)} + I_-^{(i)} \cdot I_+^{(j)}\right)\right)$. Further, since by construction $I_z$ and $\mathcal{D}$ commute, it follows $[I_{z,k}', \mathcal{D}_k'] = 0$ for all *k*. Therefore, we can separate the action of offset and dipolar couplings if we recast the one-cycle evolution operator as

$$U(t_c) = \prod_{k=1}^{n} e^{-i\tau\mathcal{H}_k'} = \prod_{k=1}^{n} Q_k e^{-i\tau\mathcal{D}_k'}\ , \tag{8}$$

where we defined $Q_k = e^{-i\omega\tau I_{z,k}'}$. This is formally identical to Eq. (3) with operators $Q_k$ effectively playing the role of pulses separated by evolution intervals $\tau$ governed by $\mathcal{D}_k'$. Therefore, we can regroup all $Q_k$ operators to recast Eq. (8) as

$$U(t_c) = \prod_{k=1}^{n} e^{-i\tau\tilde{\mathcal{D}}'_k} \prod_{k=1}^{n} Q_k \ , \tag{9}$$

where $\tilde{\mathcal{D}}'_k = \left(\prod_{l=k}^{n} Q_l\right)\mathcal{D}'_k\left(\prod_{l=k}^{n} Q_l\right)^{-1}$, i.e., an offset-dressed version of the toggling frame dipolar Hamiltonian $\mathcal{D}'_k$.

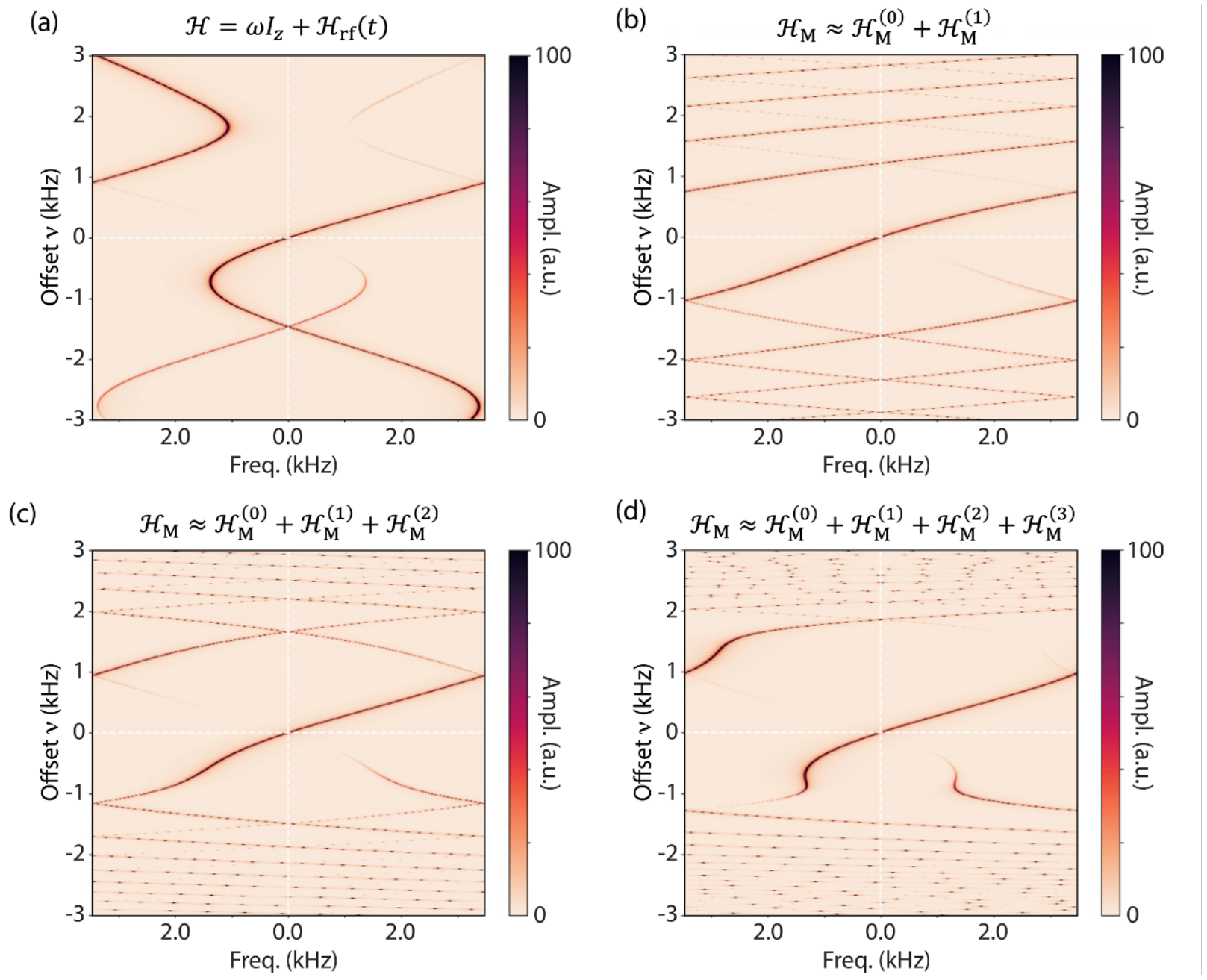

**Supplementary Figure 2 | Breakdown of AHT under detuning for the DSL-4 sequence.** Fourier spectra of magnetization trajectories originating from an initial state $\rho(0) = I_y$ under DSL-4, computed for non-interacting spins across offset frequencies $\nu$. (a) Exact simulation of the full time-dependent Hamiltonian $\mathcal{H} = \omega I_z + \mathcal{H}_{\mathrm{rf}}(t)$, showing pronounced offset-dependent features. (b–d) Approximate spectra computed using increasing orders in the Magnus expansion of $\mathcal{H}_{\mathrm{eff}}$: (b) up to first order; (c) second order, and (d) third order. All calculations exclude dipolar couplings, isolating the effects of detuning alone. Agreement between AHT and exact dynamics is restricted to a narrow region around zero offset; outside this range, even high-order AHT fails to capture the true spectral structure.

Since $Q_k$ amounts to a global rotation of angle $\omega\tau$ about the axis defined by $I'_{z,k}$, we can always write

$$\prod_{k=1}^{n} Q_k = \exp(-it_c\boldsymbol{\delta}(\omega)\cdot\mathbf{I}) \ , \tag{10}$$

with $\boldsymbol{\delta}(\omega)$ denoting an effective spin locking field acting during the cycle. Notice that $\boldsymbol{\delta}(\omega)$ depends exclusively on the chosen pulse protocol, hence creating opportunities to tailor its amplitude and orientation via sequence design, independently of system-specific parameters such as the gyromagnetic ratio or coupling strengths; we will return to this important point later (see below Section IV). On the other hand, assuming the interpulse interval is sufficiently short so that $\tau\|\tilde{\mathcal{D}}'_k\| \ll 1$, it is possible recast the effect of dipolar interactions throughout the cycle as

$$U_{\mathrm{D}}(t_c) = \prod_{k=1}^{n} e^{-i\tau\tilde{\mathcal{D}}'_k} = \exp\left(-it_c\tilde{\mathcal{D}}'_{\mathrm{M}}\right) \ , \tag{11}$$

where $\tilde{\mathcal{D}}'_{\mathrm{M}}$ denotes the Magnus dipolar Hamiltonian. Therefore, combining Eqs. (9) through (11) and making use of the Baker-Campbell-Hausdorff (BCH) formula, we write

$$U(t_c) = e^{-it_c\tilde{\mathcal{D}}'_{\mathrm{M}}} e^{-it_c\boldsymbol{\delta}(\omega)\cdot\mathbf{I}} = e^{-it_c\mathcal{H}_{\mathrm{BCH}}}, \tag{12}$$

where we defined

$$\mathcal{H}_{\mathrm{BCH}} = \delta(\omega)\cdot I_\delta + \tilde{\mathcal{D}}'_{\mathrm{M}} + \frac{t_c}{2}\left[\delta(\omega)I_\delta, \tilde{\mathcal{D}}'_{\mathrm{M}}\right] + \frac{t_c^2}{12}\left[\delta(\omega)I_\delta, \left[\delta(\omega)I_\delta, \tilde{\mathcal{D}}'_{\mathrm{M}}\right]\right] + \frac{t_c^2}{12}\left[\tilde{\mathcal{D}}'_{\mathrm{M}}, \left[\tilde{\mathcal{D}}'_{\mathrm{M}}, \delta(\omega)I_\delta\right]\right] + \cdots , \tag{13}$$

with $I_\delta$ denoting the spin operator projected in the direction parallel to $\boldsymbol{\delta}(\omega)$. Interestingly, Eq. (13) applies to all values of detuning provided the use of average Hamiltonian theory in Eq. (12) holds. Since the amplitude and tilt of $\boldsymbol{\delta}(\boldsymbol{\omega})$ can be captured through a numerical fit, Eq. (13) provides the basis for a semi-analytical solution once $\tilde{\mathcal{D}}'_{\mathrm{M}}$ is calculated to the desired degree of accuracy.

Even in the absence of explicit formulas, the comments below — presented for clarity as consecutive bullet points — help gain a qualitative understanding of the offset response presented in Fig. 2 of the main text:

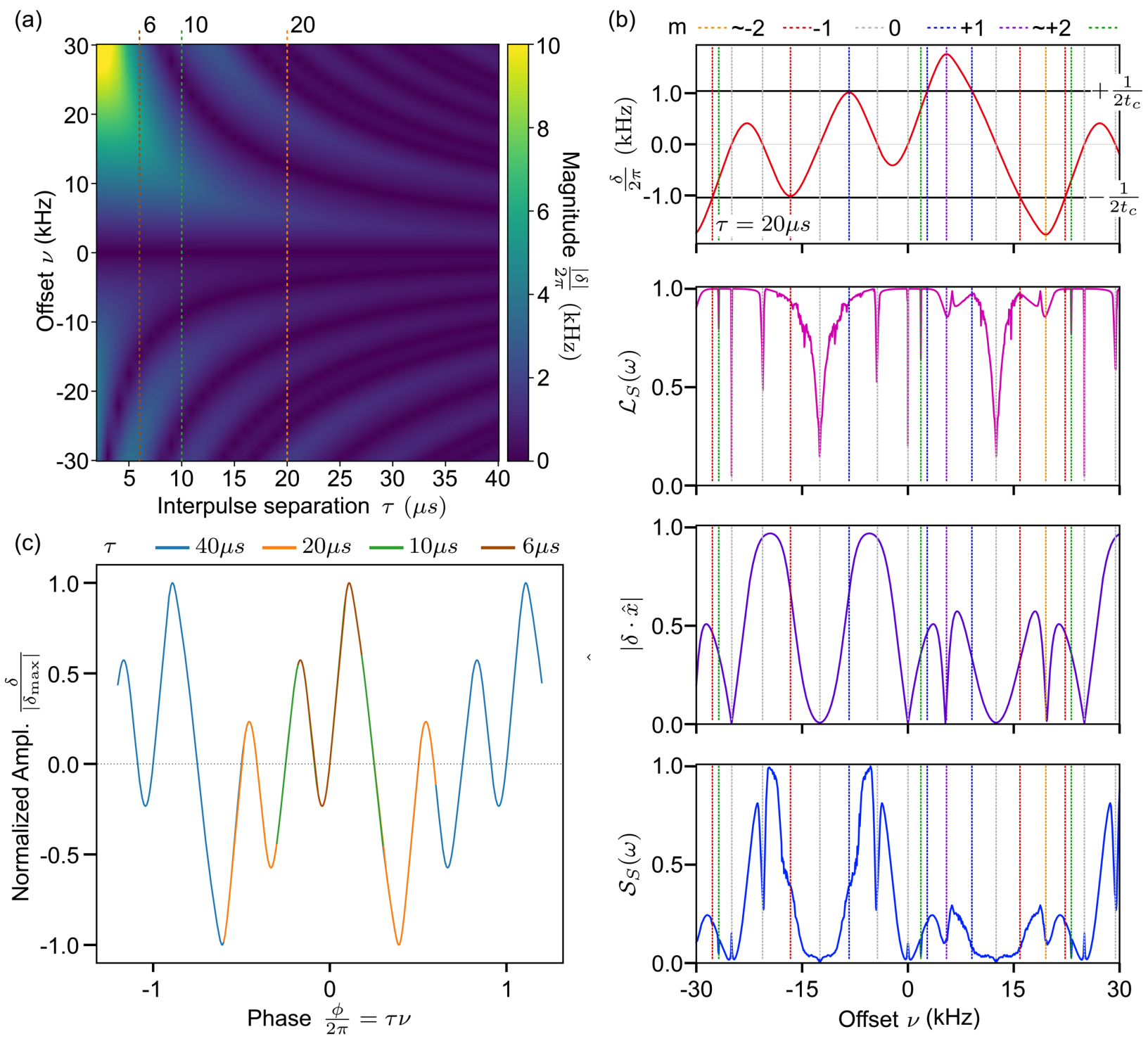

**Supplementary Figure 3 | Offset dependence of the locking field $\boldsymbol{\delta}$.** (a) Magnitude of $\boldsymbol{\delta}$ as a function of offset $\nu$ and interpulse interval $\tau$. (b) Detailed analysis for $\tau = 20$ μs: (top) signed Rabi field amplitude $\delta$, (second panel) calculated spin-locking efficiency for a 10-spin cluster with dipolar couplings representative of adamantane, (third panel) projection of $\boldsymbol{\delta}$ onto the $x$-axis (parallel to the initial magnetization), and (bottom) predicted amplitude of the normalized signal as derived from a 10-spin cluster simulation. Dashed vertical lines indicate offset frequencies where $\delta t_c = m\pi$ with integer $m$; dashed green lines mark dips of unknown origin. (c) Normalized $\delta$ amplitude in a DSL-4 protocol plotted as a function of the accumulated phase $\phi = 2\pi \times \nu\tau$ for different $\tau$ values. Colors correlate with the vertical dashed lines in (a).

- Since $e^{-i\omega\tau I'_{z,k}} = e^{-i(\omega\tau+2\pi)I'_{z,k}}$ for all values of $k$, it follows that $\delta(\omega)$ must be a periodic function of the offset with period defined by the condition $\omega_p\tau = 2\pi$; for $\tau = 20$ μs, we find $\omega_p = 2\pi \times 50$ kHz. In the same vein, for arbitrary $l, k$, it is easy to show that $Q_l \mathcal{D}'_k Q_l^{-1} = \mathcal{D}'_k$ when $\omega\tau = (2m+1)\pi$ with $m$ integer, a consequence of the bilinear nature of dipolar couplings. Therefore, for the DSL-4 protocol it follows that $\prod_{k=1}^{n} e^{-i\tau\widetilde{\mathcal{D}}'_k} = \prod_{k=1}^{n} e^{-i\tau\mathcal{D}'_k} \approx 1$ for detuning satisfying $\omega\tau = \pm\pi, \pm 2\pi, \pm 3\pi, \cdots$.
- Another case of interest is when $t_c\delta(\omega) = m\pi$, with $m$ denoting again an integer. The one-cycle evolution operator now takes the form

$$U(t_c) = U_D e^{-im\pi I_\delta} = e^{-im\pi I_\delta} U_D\,, \tag{14}$$

which holds for even and odd values of $m$ given the bilinearity of $\mathcal{D}$. Therefore, assuming a global observable $I_o$ denoting spin projection along an arbitrary direction, the detectable signal after $l$ cycles of the protocol reads

$$\langle I_o\rangle(lt_c) = Tr\left\{\left(U_D e^{-im\pi I_\delta}\right)^l \rho(0)\left(U_D e^{-im\pi I_\delta}\right)^{-l} I_o\right\} = Tr\left\{(U_D)^l \rho(0)(U_D)^{-l}\left(I_o^{\parallel} + (-1)^{lm} I_o^{\perp}\right)\right\}, \tag{15}$$

where $I_o^{\parallel}$ and $I_o^{\perp}$ respectively represent the projections of $I_o$ parallel and perpendicular to $\boldsymbol{\delta}$. Equation (15) implies that spins evolve exclusively under the action of $\widetilde{\mathcal{D}}'_M$ and thus decohere over time, leading to a dip in the amplitude of the spin-locked signal. In particular, this must be the case for $m = 0, \pm 1$, meaning one should expect signal dips when $\delta(\omega)$ vanishes or when $\delta(\omega)/2\pi = \pm 1/2t_c$. Notice that the same argument applies to the case $\omega\tau = m\pi$, except that here the decay is slower since, under these conditions, $\widetilde{\mathcal{D}}'_M = \mathcal{D}'_M \approx 0$ in a DSL-4 sequence.

We summarize the above observations in Supplementary Fig. 3, where we first calculate the locking field amplitude in DSL-4 as a function of detuning and interpulse interval (Supplementary Fig. 3a). Focusing on the case $\tau = 20$ μs, we carry out a detailed analysis in Supplementary Fig. 3b where we first reproduce the offset dependence of $\delta$, and then plot the locking efficiency $\mathcal{L}_S(\omega)$ as extracted from simulating the dynamics of a 10-spin cluster used as a proxy for adamantane (top and second plots, respectively); we simulate the system response in the bottom panel once we calculate the projection of the initial magnetization ($x$ in this case) along $\boldsymbol{\delta}$ (third panel from top). As noted above, we find $\delta(\omega)$ to be periodic with a period given by $\nu_p = \tau^{-1} = 50$ kHz; further, we find that most deeps in $\mathcal{L}_S(\omega)$ — and correspondingly, in the ensuing simulated signal $\mathcal{S}_S(\omega)$ — can be traced back to $t_c\delta(\omega) = m\pi$ for different values of $m$ (color-coded vertical lines). Lastly, we note that since all factors in the operator product $\prod_{k=1}^{n} Q_k$ simultaneously depend on the detuning and interpulse separation through the product $\omega\tau$, the locking field can be expressed as a function of the phase $\phi = \omega\tau$ whose exact shape depends only the chosen pulse protocol (DSL-4 in Supplementary Fig. 3c).

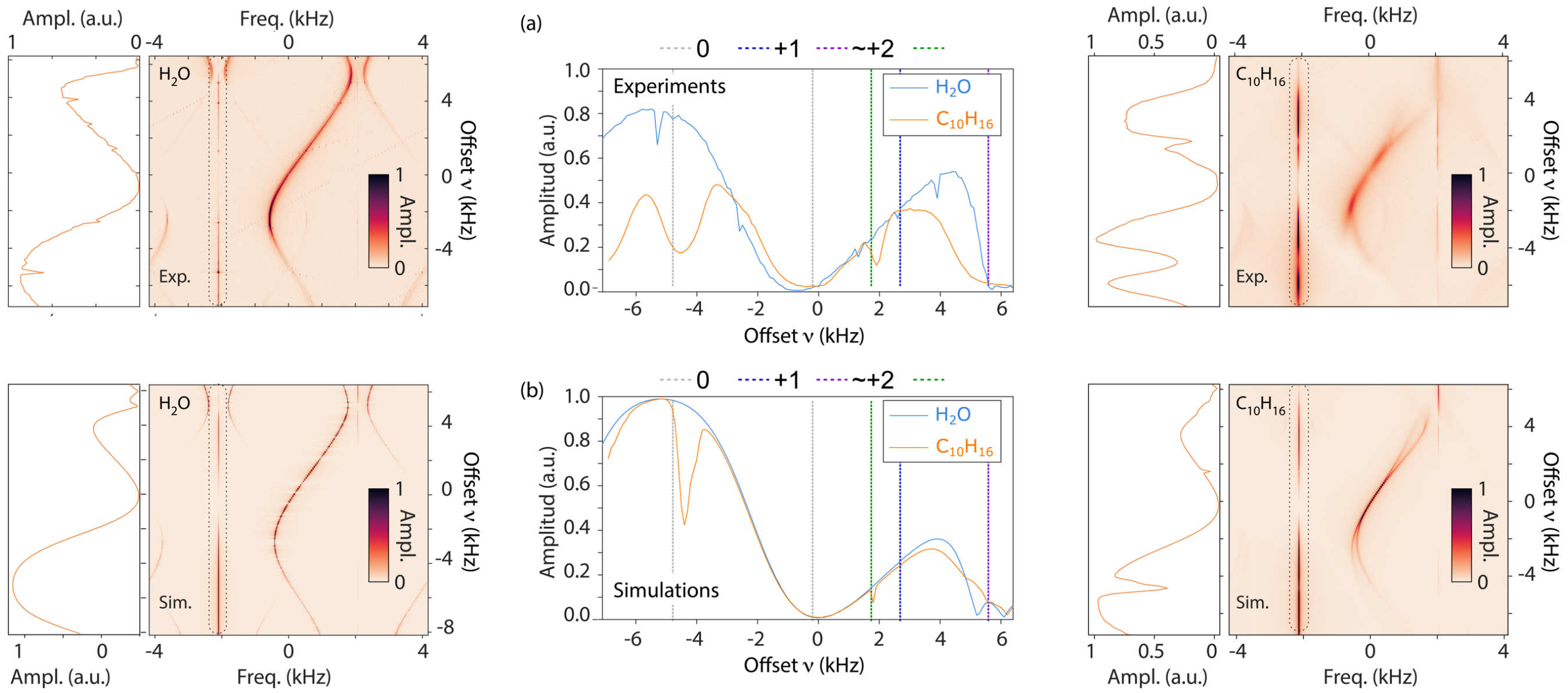

**Supplementary Figure 4 | Experimental and simulated spectral response under DSL-4 control for interacting and non-interacting systems.** (a) Experimental results acquired using the low-field setup. (Center panel) Intensity of the Fourier amplitude at $-1/t_c^{\text{DSL}}$ as a function of frequency offset for both water (blue) and adamantane (orange); vertical dashed lines indicate offset frequencies where $\delta t_c = m\pi$, following the nomenclature in Supplementary Fig. 3. (Left panel) Full two-dimensional short-time Fourier transform (STFT) of the water signal under DSL-4 control across a range of detunings. (Right panel) Same as left, but for adamantane. Compared to water, adamantane shows suppressed spectral weight at select detunings, indicative of decoherence due to internal dipolar interactions. (b) Same as in (a) but as derived from numerical simulations of a 10-spin cluster with or without dipolar interactions, here included for ease in comparison. Throughout these experiments, the interpulse separation is $\tau = 20$ μs and the $\pi/2$-pulse duration is 3.6 μs.

## II. Extended Experiments

This section presents additional information on the experimental methods and supporting data used to investigate the emergence of dynamic spin locking under DSL-4 control. Our study relies on measurements from two distinct NMR setups: A low-field benchtop system (Bruker Minispec) operating at 20 MHz proton Larmor frequency (0.5 T), and a high-field spectrometer (Bruker Avance III HD, 750 MHz proton frequency corresponding to a 17.6 T superconducting magnet) equipped with multichannel rf control.

The interacting spin system studied here is adamantane ($C_{10}H_{16}$), a prototypical organic solid in which rapid on-site molecular tumbling averages out intramolecular dipolar couplings. Notably, two forms of adamantane were employed: A relatively dry crystalline form used in the high-field setup, and a more hydrated form in the low-field system that exhibited reduced dipolar couplings (respectively, 14 and 5.5 kHz linewidths). As a non-interacting reference, we used liquid water, which lacks substantial proton–proton dipolar couplings due to rapid molecular motion.

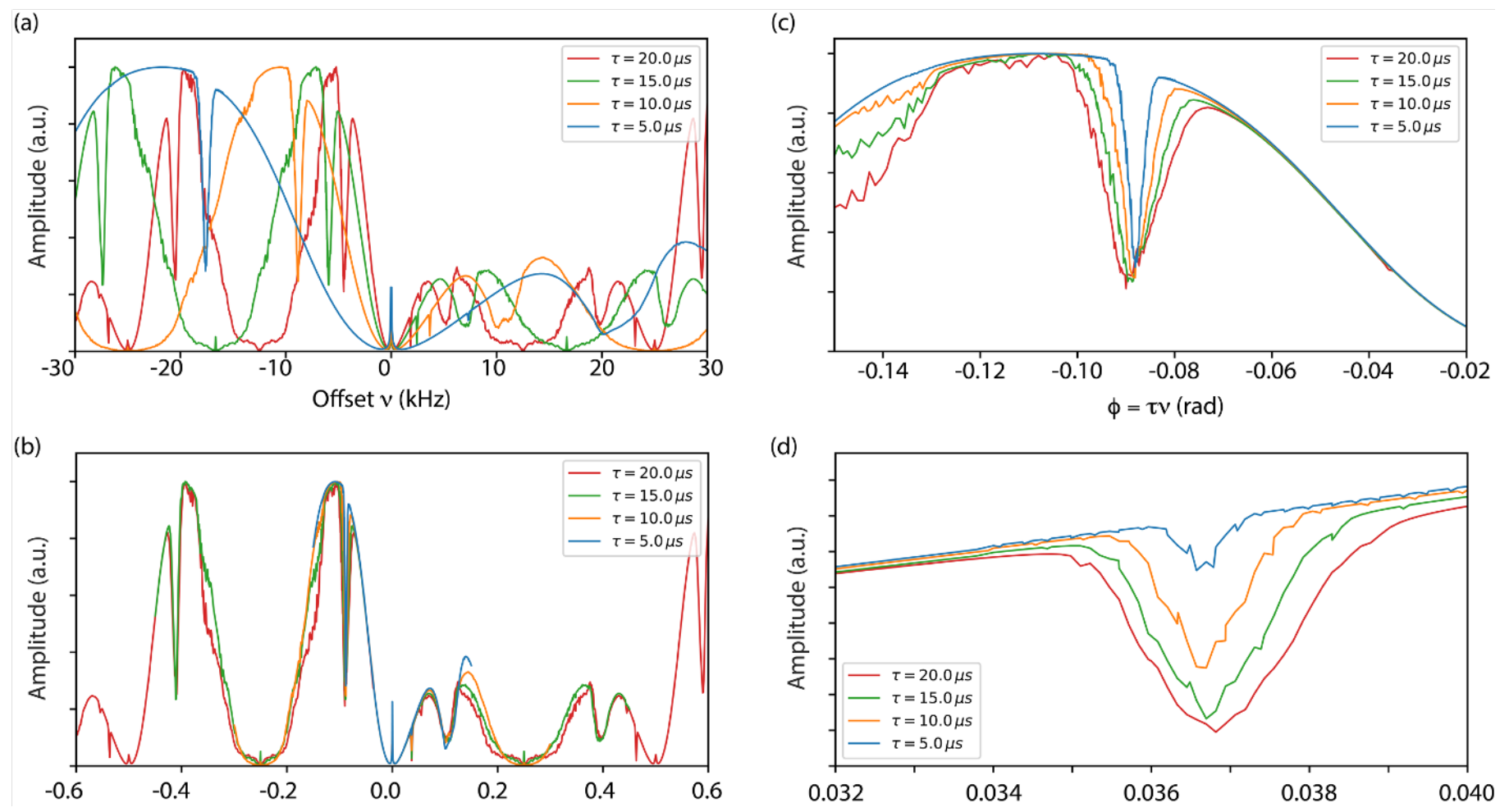

**Supplementary Figure 5 | Offset-normalized signal response in a 10-spin interacting cluster under DSL-4 control.** (a) Calculated signal amplitudes as a function of offset ν for different interpulse delays $\tau$. (b) Same data as in (a), but plotted against the dimensionless phase variable $\phi = \tau\nu$, revealing a near-universal structure that collapses the curves for different τ values onto a common profile. (c-d) Zoom-in views of selected features in (b), highlighting sharp spectral dips that recur at well-defined values of $\phi$.

To isolate the contribution of internal dipolar interactions from that of the pulse protocol itself, we performed both comparative experiments and supporting simulations. In particular, we contrasted the behavior of interacting and non-interacting proton ensembles across a range of frequency offsets. As shown in Supplementary Fig. 4, we monitored the spectral response of water and adamantane under DSL-4 sequences using short-time Fourier analysis. While both systems exhibit well-defined offset-dependent spectral features, only adamantane — where residual dipolar interactions persist — shows suppression of signal amplitude at specific detunings, consistent with interaction-induced decoherence. Simulations of dipolar spin clusters reproduce these trends and confirm that the observed spectral structure is a fingerprint of underlying spin-spin interactions. On the other hand, we find a slight disagreement between the observed and predicted positions of the first possitive-offset dip (green and blue dashed vertical lines, respectively); we presently ignore the reasons for this discrepancy.

To better characterize the structure of these dips, we numerically examined how the response depends on the pulse timing parameter $\tau$. As shown in Supplementary Fig. 5, we calculated the locked signal amplitude for a 10-spin cluster as a function of detuning and for various $\tau$ values. Panel (a) shows the raw signal as a function of offset $\nu$, which reveals a rapidly evolving landscape of dips and bands. In Panel (b), we replot the same data as a function of the dimensionless phase $\phi = \tau\nu$, revealing a collapse of the response curves onto a common pattern. This indicates that the dip structure is governed not by the absolute detuning but by the dynamical phase accumulated between pulses, consistent with the formalism in Section I. Panels (c) and (d) zoom in on the first negative- and positive-offset dips, emphasizing their reproducibility but also highlighting their dependance on the interpulse separation.

The exponential decay in the NMR signal envelope — comparable but faster than the adamantane spin-lattice relaxation time, ~1 s — prompted us to explore possible sources of decoherence not captured in our numerical modeling. One first hypothesis is that the decay arises from heteronuclear dipolar interactions with nearby $^{13}C$ nuclei, whose dynamics are not controlled by the DSL-4 sequence. Because the Minispec system lacks a second RF channel for $^{13}C$ decoupling, we could not test this hypothesis directly on that platform. This limitation motivated a second set of experiments using a high-field spectrometer (750 MHz proton frequency), which offers full heteronuclear decoupling capabilities and tighter control over pulse parameters.

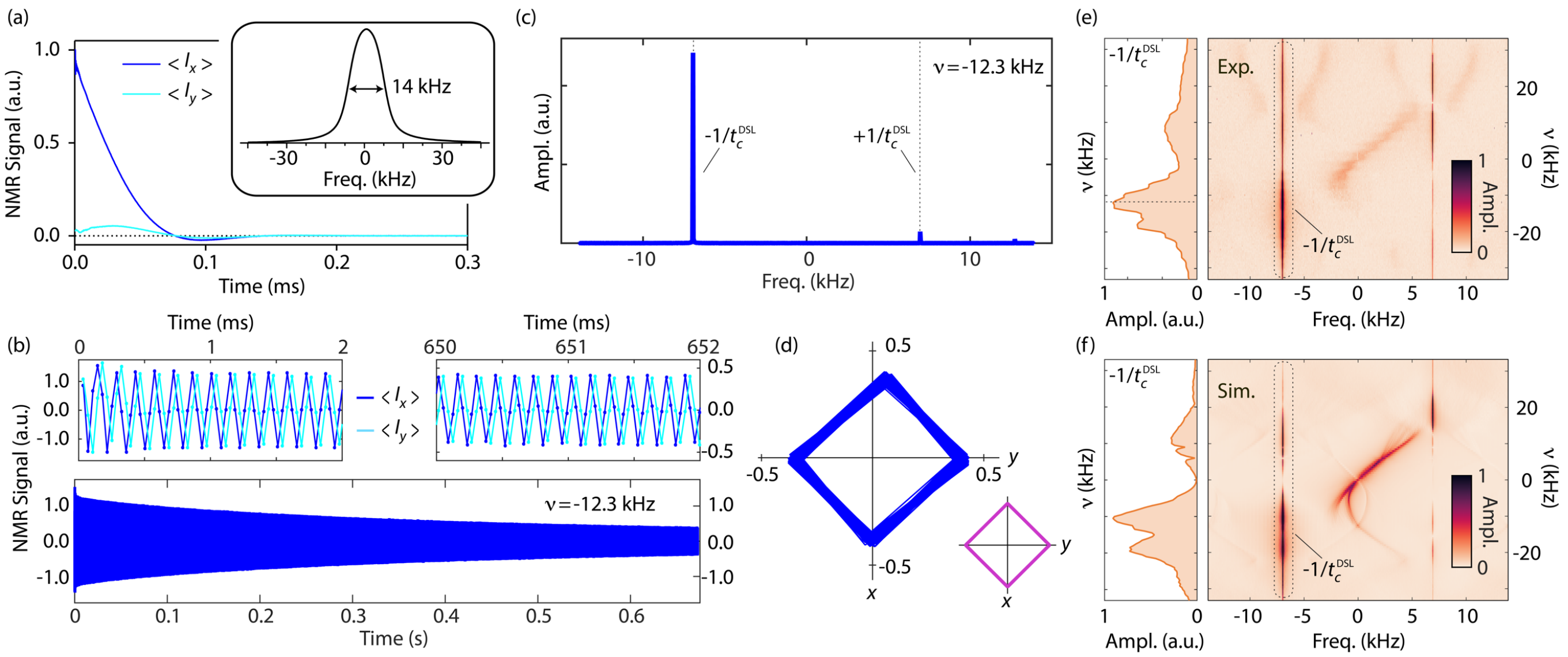

**Supplementary Figure 6 | High-field response of adamantane under DSL-4 control.** (a) Free induction decay (FID) of the high-field adamantane sample acquired on the 750 MHz spectrometer, showing rapid signal decay due to strong dipolar broadening. The corresponding spectrum (inset) reveals a linewidth of ~14 kHz, significantly broader than that observed in the low-field Minispec system, consistent with enhanced intramolecular couplings in this drier, crystalline form of adamantane. (b) Time-domain signal under DSL-4 control with $\pi/2$-pulse duration of 1.975 μs and interpulse separation $\tau = 6$ μs, recorded at detuning $\nu = -12.3$ kHz. The locked magnetization persists for over 600 ms, substantially longer than achievable in the low-field setup but still shorter than the time defined by spin-lattice relaxation. Insets show zoomed-in views of the oscillatory dynamics at early and late times. (c) Fourier transform of the signal in (b), showing a sharp spectral feature centered at $-1/t_c^{\mathrm{DSL}}$, corresponding to the effective spin-locking frequency. (d) Locked magnetization trajectory reconstructed from the measured signal (blue) and compared with the ideal theoretical orbit (magenta). (e) Offset-dependent spectral response under DSL-4 control measured for this sample. (f) Simulated response of a dipolar-coupled 10-spin cluster under the same experimental conditions, reproducing the observed offset dependence and spectral structure.

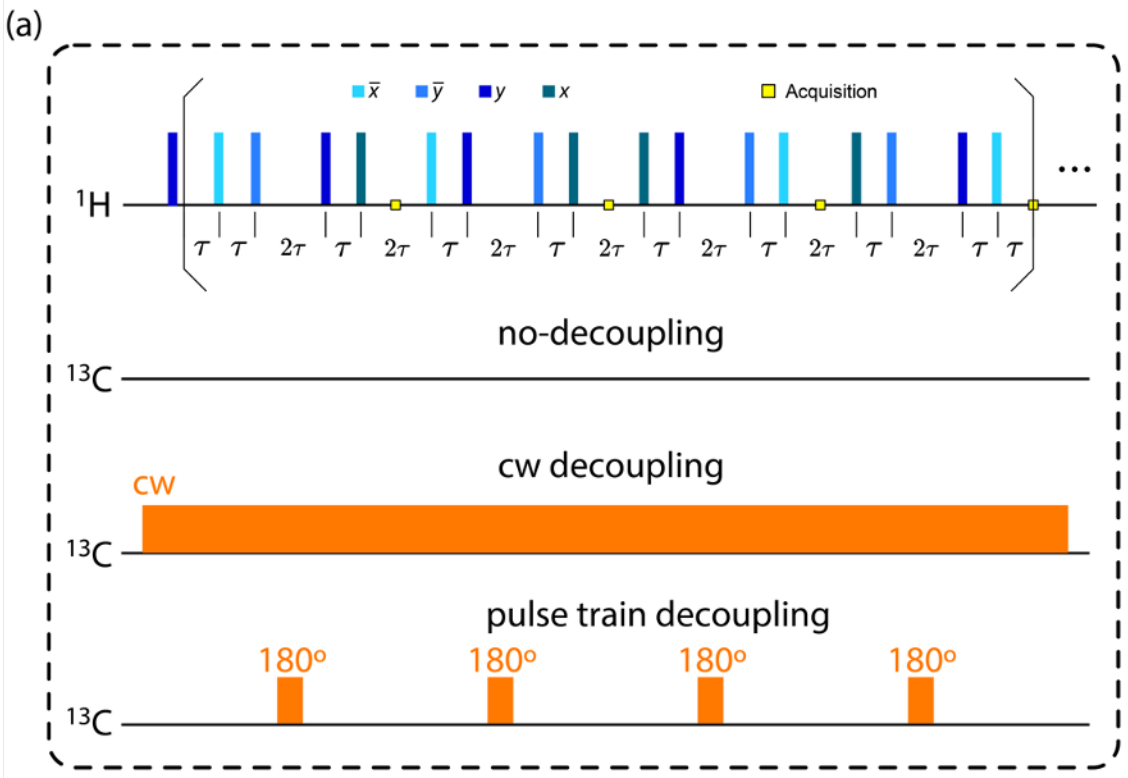

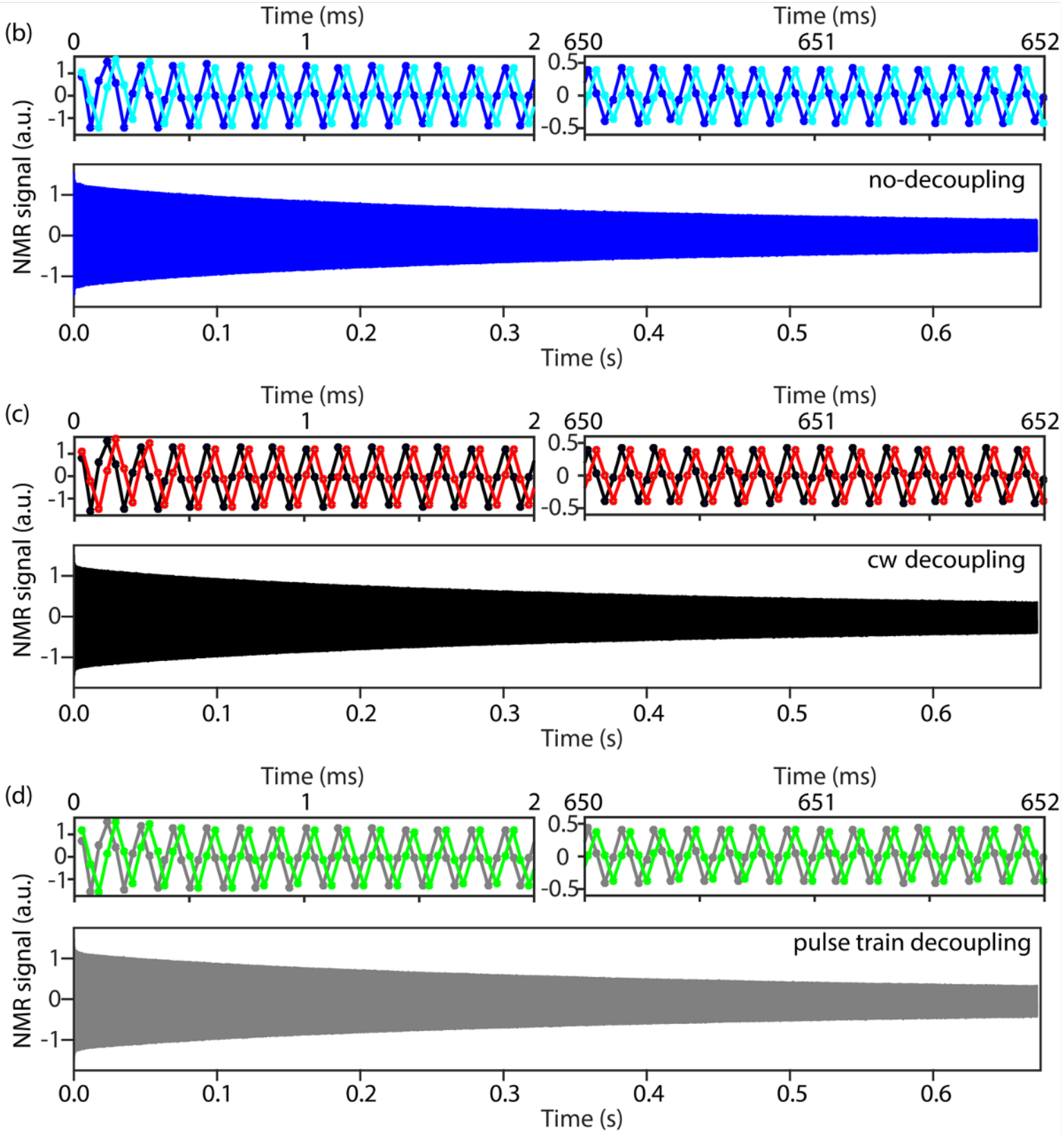

**Supplementary Figure 7 | Effect of heteronuclear ($^{13}$C) decoupling on dynamical spin locking in proton ensembles.** (a) Schematic representation of the experimental pulse sequences applied to the $^1$H spins, including DSL-4 control (top line) and three different $^{13}$C decoupling conditions: no decoupling (second line), continuous wave (cw) decoupling resonant with the $^{13}$C frequency (third line), and synchronized $^{13}$C π-pulse trains intercalated within the DSL-4 cycle (bottom line). (b) Observed DSL-4 spin response under optimal offset ($\nu = -18$ kHz for an interpulse interval $\tau = 6$ μs) without any $^{13}$C decoupling. The main panel shows long-lived dynamical spin-locking behavior over hundreds of milliseconds, while the two insets present zoomed views of the oscillatory signal at the beginning and end of the acquisition window, demonstrating persistence of coherent dynamics. (c) Same as in (b), but with cw $^{13}$C decoupling. (d) Same as in (b), but using synchronized pulse-train $^{13}$C decoupling. The similarity across all three conditions confirms that $^{13}$C–$^1$H coupling is not the dominant source of the observed decay, validating that the locked magnetization trajectories are not limited by heteronuclear dephasing.

The experiments on the high-field system were designed to approximately replicate the DSL-4 spin-locking conditions observed in the Minispec, while enabling the selective decoupling of $^{13}$C spins; one notable difference, however, was the effective dipolar coupling in the adamantane sample we used (featuring a linewidth of 14 kHz, substantially greater than the 5.5 kHz linewidth observed in the Minispec sample). The DSL-4 sequence used a 90º pulse duration of 1.975 μs (a nutation frequency of 126.6 kHz), a minimal interpulse spacing of 6 μs, and consisted of 75,000 π/2 pulses (equivalent to 4688 full DSL-4 cycles). The detection dead time was 4.75 μs, and all measurements were performed using single-scan acquisition at room temperature. We summarize these results in Supplementary Fig. 6, where we also include simulations under the high-field conditions (in good agreement with our observations).

We next tested whether these observed decay profiles are influenced by heteronuclear interactions with $^{13}$C spins. The high-field platform enables direct experimental evaluation of this question by introducing $^{13}$C decoupling schemes. We tested three conditions: (*i*) no $^{13}$C decoupling, (*ii*) continuous-wave (cw) decoupling at a $^{13}$C nutation frequency of 25 kHz, and (*iii*) a synchronized pulse train decoupling scheme using 180º $^{13}$C pulses of 10 μs duration (nutation frequency of 50 kHz). The experimental protocol is schematically depicted in Supplementary Fig. 7a, and the corresponding results are shown in Supplementary Figs. 7b through 7d.

Remarkably, all three decoupling schemes yielded nearly identical DSL-4 responses, with long-lived periodic magnetization and indistinguishable decay envelopes. These results indicate that interactions with $^{13}$C spins — whether refocused or not — are not responsible for the decay observed in either the Minispec or high-field Bruker platforms. This suggests that the origin of the observed decay must lie in other mechanisms not captured by the idealized spin models used in simulation, and motivates future efforts to identify and characterize these sources.

### III. Spin Thermalization and Entropy Flow under DSL Control

To probe the dynamical behavior during the control protocol, we performed simulations starting from non-equilibrium initial states. Supplementary Figure 8 illustrates the dynamics under DSL-4 control for two different cases. In Panel 8a, we examine the evolution of magnetization when only a single spin (spin 0) is initially polarized. The top subpanel tracks the time evolution of the transverse magnetization $\langle I_{x,0} \rangle$ of this spin (red trace), as well as the averaged transverse magnetization of the remaining nine spins (light blue trace). While $\langle I_{x,0} \rangle$ exhibits a gradual decay, a complementary growth in the surrounding spins is observed, signaling polarization transfer throughout the cluster. Notably, the total transverse magnetization $\langle I_x \rangle$ (bottom subpanel) remains nearly constant in both amplitude and phase structure, despite the internal redistribution. This internal re-equilibration process can be understood as a form of dynamical spin thermalization driven by dipolar interactions in the dressed frame defined by the DSL protocol. In this frame, the magnetization is effectively stabilized, yet remains susceptible to internal interactions that facilitate entropy flow and polarization exchange. The contrast with the non-interacting case (Supplementary Fig. 8b) — in which the initially polarized spin retains its magnetization indefinitely while the others remain unpolarized — underscores the essential role of dipolar couplings in enabling thermalization. It also highlights a potential limitation of ensemble observables: While the total signal $\langle I_x \rangle$ appears similar in both cases, the underlying microscopic behavior is fundamentally different.

This behavior bears a conceptual resemblance to traditional spin-locking experiments in solid-state NMR. There, the application of a strong continuous RF field redefines the quantization axis, aligning it along the transverse direction (e.g., $x$), such that magnetization initialized along this axis becomes effectively longitudinal. Residual dipolar interactions in the rotating frame can then mediate spin flip-flops, enabling polarization to spread across the sample, a process often described as spin diffusion under spin-lock. In our case, a similar redistribution of polarization is observed despite the fact that, under dynamical spin locking, the magnetization only transiently aligns with the effective quantization axis.

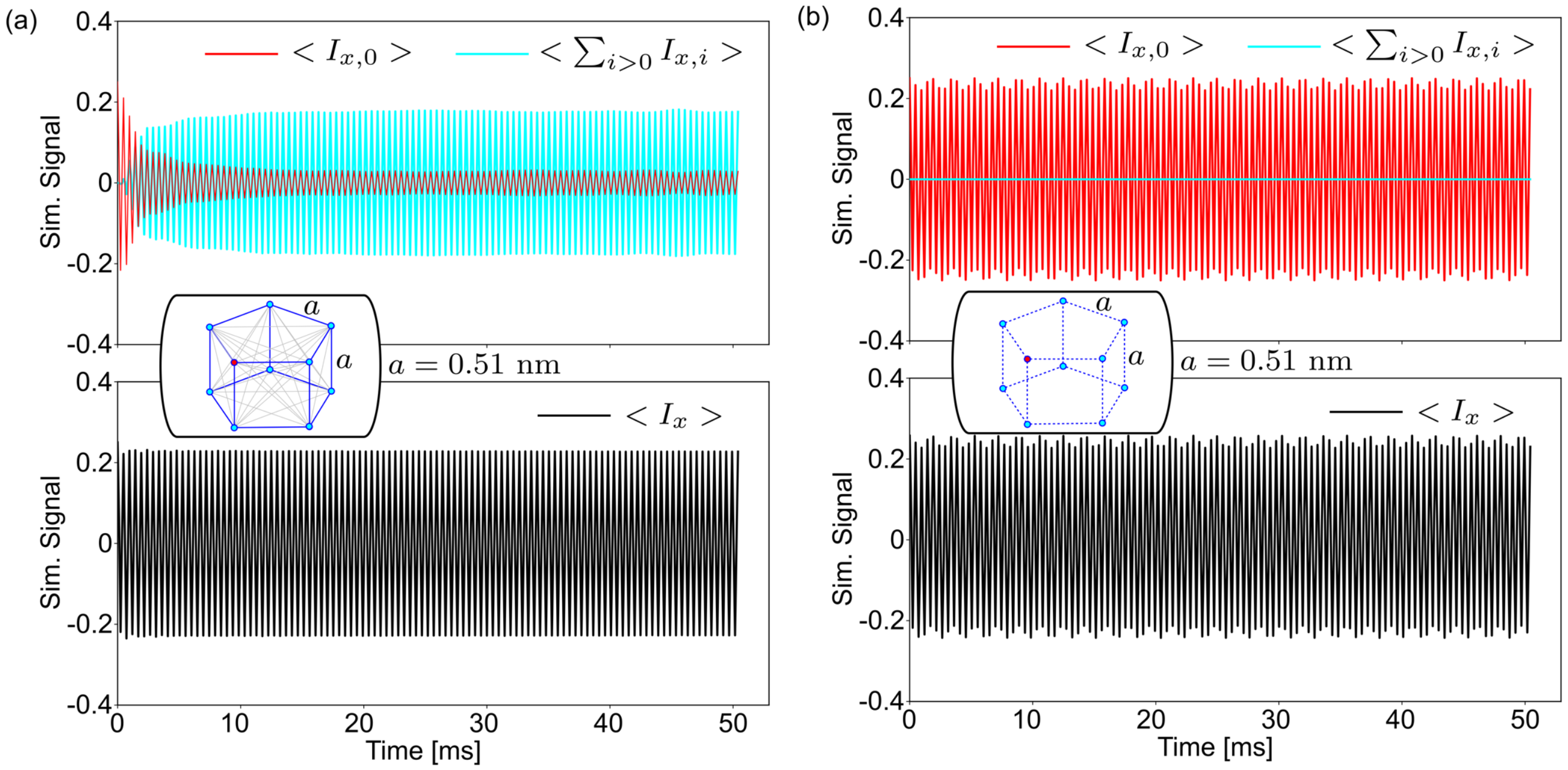

**Supplementary Fig. 8 | Role of initial spin polarization and interactions in DSL-4.** (a) Simulated spin dynamics of a 10-proton cluster with dipolar interactions arranged in the geometry of two stacked pentagons, subject to DSL-4 control ($\tau = 20$ μs, $\nu = 5.8$ kHz). Only one spin (spin 0) is initially polarized. The top panel shows the evolution of the transverse magnetization of the initially polarized spin ($\langle I_{x,0} \rangle$, red trace) and the average magnetization of the remaining spins in the cluster ($\langle I_{x,i} \rangle$ for $i > 0$, light blue). The transfer of polarization from the seed spin to the rest of the cluster reflects the development of collective dynamics enabled by dipolar interactions. The bottom panel shows the total transverse magnetization $\langle I_x \rangle$, which combines the contributions of all spins. Its nearly constant amplitude indicates that the total magnetization is preserved while being redistributed. (b) Identical protocol and initial condition, but with all dipolar couplings turned off, i.e., a non-interacting spin ensemble. The top panel shows that only the initially polarized spin evolves coherently, while the remaining spins stay at zero polarization throughout. The bottom panel displays the total transverse magnetization $\langle I_x \rangle$, which still shows an oscillatory pattern due to the single evolving spin.

This feature highlights the rich structure of the driven system and supports the interpretation of DSL-4 as a form of dynamic frame stabilization compatible with spin transport. More broadly, this opens the possibility of using dynamical spin locking to engineer polarization transfer protocols across heteronuclear spin species — akin to Hartmann–Hahn matching conditions — or between electron and nuclear spins, as exploited in NOVEL and other dynamic nuclear polarization schemes.

Specifically, the double-rotating frame Hamiltonian for a system containing dipolar-coupled spins $I$ and $S$ takes the form

$$\mathcal{H}(t) = \omega_I I_z + \omega_S S_z + \mathcal{D}_I + \mathcal{D}_S + \mathcal{H}_{\mathrm{rf}}^I(t) + \mathcal{H}_{\mathrm{rf}}^S(t) + \sum_{i>j} 2\pi J_{ij}\, I_z^{(i)} S_z^{(j)}, \tag{16}$$

where, as before, $\omega_{I,S}$ denotes the offset frequencies, $\mathcal{D}_{I,S}$ is the homo-spin dipolar coupling Hamiltonian for each spin species, $\mathcal{H}_{\mathrm{rf}}^{I,S}$ captures the (time-dependent) radio-frequency control pulses, and the last term is the hetero-spin coupling. Note that $\mathcal{H}(t)$ is a piecewise constant Hamiltonian. Following the formalism in Section I, we write the evolution operator over one cycle, $U(t_{\mathrm{c}})$, as

$$U(t_{\mathrm{c}}) = \prod_{k=1}^{n} P_k e^{-i\tau\mathcal{H}} = \prod_{k=1}^{n} e^{-i\tau\mathcal{H}_k'} \prod_{k=1}^{n} P_k = \prod_{k=1}^{n} e^{-i\tau\mathcal{H}_k'}\,, \tag{17}$$

where operators $P_k$ denote rf pulses, $\mathcal{H}_k' = (\prod_{l=k}^{n} P_l)\mathcal{H}(\prod_{l=k}^{n} P_l)^{-1}$ is the toggling frame Hamiltonian during the $k$-th free evolution interval, and we have used $\prod_{k=1}^{n} P_k = 1$ for DSL-4. Note that $P_k = P_k^I P_k^S$, since the rf Hamiltonians act separately on each spin species. More importantly, since we apply both sequences synchronously, the toggling frame is shared for both $I$ and $S$ spins. The toggling frame Hamiltonian $\mathcal{H}_k'$ results from transforming both $I$ and $S$ operators simultaneously

$$\mathcal{H}_k' = \omega_I I_{z,k}' + \omega_S S_{z,k}' + \mathcal{D}_{I,k}' + \mathcal{D}_{S,k}' + \mathcal{J}_{IS}'\,, \tag{18}$$

where we introduced the notation $\mathcal{J}_{IS}' = \sum_{i>j} 2\pi J_{ij} I_{z,k}^{(i)'} S_{z,k}^{(j)'}$ to denote the hetero-spin coupling in the double toggling frame. Since $\omega_I I_{z,k}' + \omega_S S_{z,k}'$ commutes with every dipolar coupling term, we can separate the action of free evolution under offset to recast the one-cycle evolution operator as

$$U(t_{\mathrm{c}}) = \prod_{k=1}^{n} e^{-i\tau\mathcal{H}_k'} = \prod_{k=1}^{n} e^{-i\tau(\widetilde{\mathcal{D}}_{I,k}' + \widetilde{\mathcal{D}}_{S,k}' + \tilde{\mathcal{J}}_{IS,k}')} \prod_{k=1}^{n} Q_k\,, \tag{19}$$

where we defined $Q_k = e^{-i\tau(\omega_I I_k' + \omega_S S_k')}$, $\widetilde{\mathcal{D}}_{I,S,k}' = (\prod_{l=k}^{n} Q_l)\mathcal{D}_{I,S,k}'(\prod_{l=k}^{n} Q_l)^{-1}$, and $\tilde{\mathcal{J}}_{IS,k}' = (\prod_{l=k}^{n} Q_l)\mathcal{J}_{IS,k}'(\prod_{l=k}^{n} Q_l)^{-1}$ (see Eqs. (8) and (9) above). Furthermore, we can always write

$$\prod_{k=1}^{n} Q_k = \exp(-it_c(\boldsymbol{\delta}_I(\omega_I)\cdot\mathbf{I} + \boldsymbol{\delta}_S(\omega_S)\cdot\mathbf{S}))\,. \tag{20}$$

Following Section I, we assume the interpulse interval is sufficiently short so that $\tau\|\widetilde{\mathcal{D}}_k'\|, \tau\|\tilde{\mathcal{J}}_k'\| \ll 1$, and therefore it is possible to recast the effect of dipolar interactions throughout the cycle as

$$U_{\mathrm{D}}(t_c) = \prod_{k=1}^{n} e^{-i\tau(\widetilde{\mathcal{D}}_{I,k}' + \widetilde{\mathcal{D}}_{S,k}' + \tilde{\mathcal{J}}_{IS,k}')} = \exp\left(-it_c\left(\widetilde{\mathcal{D}}_{I,\mathrm{M}}' + \widetilde{\mathcal{D}}_{S,\mathrm{M}}' + \tilde{\mathcal{J}}_{IS,\mathrm{M}}'\right)\right), \tag{21}$$

where the subscript M denotes the Magnus dipolar Hamiltonian. Finally, making use of the Baker-Campbell-Hausdorff (BCH) formula, we write

$$U(t_c) = e^{-it_c(\widetilde{\mathcal{D}}_{I,\mathrm{M}}' + \widetilde{\mathcal{D}}_{S,\mathrm{M}}' + \tilde{\mathcal{J}}_{IS,\mathrm{M}}')} \mathrm{e}^{-it_c(\boldsymbol{\delta}_I(\omega_I)\cdot\mathbf{I} + \boldsymbol{\delta}_S(\omega_S)\cdot\mathbf{S})} = e^{-it_c\mathcal{H}_{\mathrm{BCH}}},$$

where, in analogy to Eq. (13), we defined

$$\mathcal{H}_{\mathrm{BCH}} = \boldsymbol{\delta}_I(\omega_I)\cdot\mathbf{I} + \boldsymbol{\delta}_S(\omega_S)\cdot\mathbf{S} + \widetilde{\mathcal{D}}_{I,\mathrm{M}}' + \widetilde{\mathcal{D}}_{S,\mathrm{M}}' + \tilde{\mathcal{J}}_{IS,\mathrm{M}}' + \cdots. \tag{22}$$

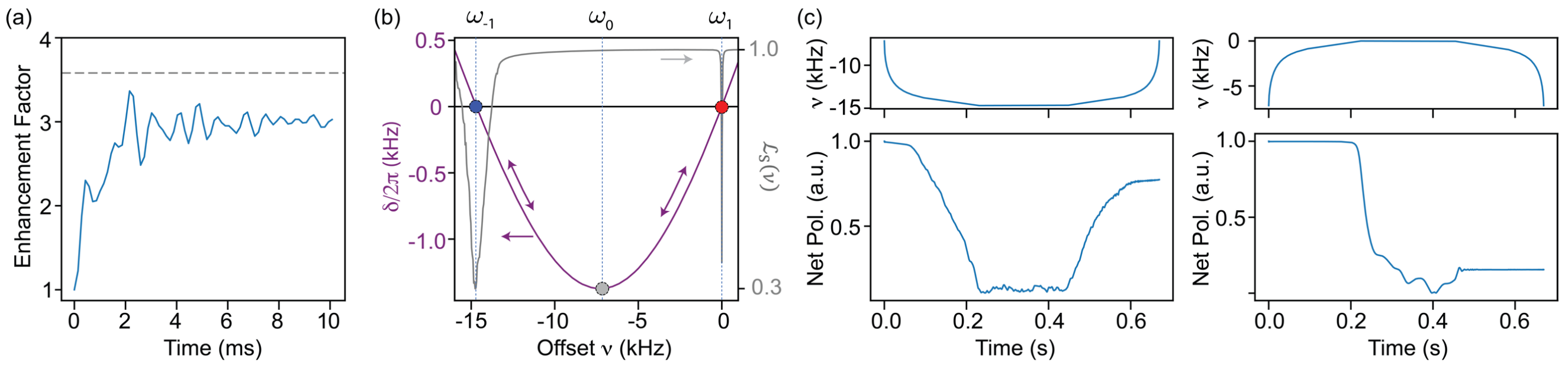

**Supplementary Figure 9 | Offset-engineered polarization transfer and adiabatic demagnetization under dynamic spin locking.** (a) Offset engineered Hartmann–Hahn polarization transfer in a cluster comprising nine $^1H$ and one $^{13}C$ spins. This protocol uses synchronous DSL-4 sequences on the $^1H$ and $^{13}C$ spins featuring identical timing and relative offset; we observe rapid polarization transfer even though the $^1H$ and $^{13}C$ rf amplitudes (respectively, 125 kHz and 62.5 kHz) do not match the Harmann-Hahn condition. The limit polarization falls slightly below the predicted maximum (dashed horizontal line), a behavior we attribute to the finite size of the spin cluster. (b) Calculated locking field δ as a function of offset assuming an interpulse separation $\tau = 6$ µs. Blue and red dots mark select roots of δ with distinct properties: The Magnus dipolar Hamiltonian remains finite at the former, while it vanishes at the latter, a behavior reflected in the locking efficiency $\mathcal{L}_S(\nu)$ (faint trace in the back). (c) Simulations of adiabatic demagnetization in the rotating frame. Modulating the offset between $\omega_0$ and $\omega_{-1}$ (respectively, grey and blue dots in (a)) allows reversible Zeeman–dipolar exchange (left), while cycling the detuning between $\omega_0$ and $\omega_1$ (red dot in (a)) leads to signal loss due to vanishing dipolar order (right).

The term $\tilde{\mathcal{J}}'_{IS,\mathrm{M}}$ captures the effective coupling between the $I$ and $S$ spins, and is therefore responsible for the exchange of polarization. To gain intuition on why this transfer takes place in general, it is useful to consider the simplified case of only two spins (which, for simplicity, we will refer to as $I$ and $S$). In this case, Eq. (22) simplifies to

$$\mathcal{H}_{\mathrm{BCH}} = \boldsymbol{\delta}_I(\omega_I)\cdot\mathbf{I} + \boldsymbol{\delta}_S(\omega_S)\cdot\mathbf{S} + 2\pi J \tilde{I}'_{z,\mathrm{M}}\tilde{S}'_{z,\mathrm{M}} + \cdots. \quad (23)$$

In a situation where $\delta_I(\omega_I), \delta_S(\omega_S) > 2\pi J$, Eq. (23) describes the equivalent of a double resonance experiment; we expect polarization transfer when (i) $\delta_I(\omega_I) = \delta_S(\omega_S)$ (the equivalent of the Hartmann-Hahn condition), and (ii) $\left[\boldsymbol{\delta}_I(\omega_I)\cdot\mathbf{I}, \tilde{I}'_{z,\mathrm{M}}\right]$ and $\left[\boldsymbol{\delta}_S(\omega_S)\cdot\mathbf{S}, \tilde{S}'_{z,\mathrm{M}}\right]$ are both non null (in which case a flip-flop term emerges).

Supplementary Fig. 9 recaptures the experiments in Fig. 3 of the main text, this time through numerical simulations of the spin dynamics in engineered spin clusters. Panel 9a showcases a complementary application, namely, heteronuclear polarization transfer from 1H to 13C using DSL-4 sequences applied synchronously to both spin species. The protocol sets identical pulse timings and relative offsets for 1H and 13C channels, but does not rely on matching nutation frequencies. Despite the absence of traditional Hartmann–Hahn matching (the rf amplitudes differ by a factor of two), we observe efficient polarization exchange between the two species. This robustness arises from offset engineering: When both species are driven at locking-compatible detunings, they share a common rotating frame structure where inter-species dipolar terms can act effectively. In this sense, DSL-4 induces a dynamically matched frame without requiring rf amplitude alignment, enabling transfer across large gyromagnetic ratios and simplifying rf hardware constraints.

Panels 9b and 9c consider the system spin dynamics as we implement adiabatic demagnetization in the rotating frame (ADRF). In ADRF, one slowly varies the offset $\nu = \omega/2\pi$, sweeping across regions of finite dipolar coupling to transfer magnetization from Zeeman to dipolar order. Supplementary Fig. 9c (left panels) shows this process numerically: By modulating the offset between two locking roots ($\omega_0$ and $\omega_{-1}$), we enable reversible magnetization cycling. The total polarization follows the detuning trajectory, recovering most of its amplitude at the end of the sweep. In contrast, cycling the offset through a magnetically inert root (right panels, $\omega_0$ to $\omega_1$) results in irreversible signal loss, consistent with the vanishing of the offset-dressed dipolar coupling in that regime. These results provide microscopic validation of the δ-structured picture: Effective locking can persist throughout the ADRF sweep only if dipolar-mediated redistribution remains active. Partial recovery stems from finite sweep speeds and could be improved via slower ramps or tailored trajectories.

These results highlight the broader utility of DSL protocols for actively controlling entropy flow, enabling thermalization, polarization redistribution, and spin transport — all within a single unified dynamical framework. As such, DSL sequences offer a flexible alternative to conventional spin-locking techniques, opening new avenues for coherent control and quantum thermodynamics in interacting spin systems.

## IV. Artificial-intelligence-derived control protocols

Even when the microscopic Hamiltonian of a quantum many-body system is known, it is often nontrivial — or outright intractable — to determine a pulse sequence that drives the system toward a desired effective Hamiltonian. This challenge becomes especially pronounced in systems with complex internal interactions, large Hilbert space dimensions, or limited access to intuitive design heuristics. To overcome these limitations, we turn to a data-driven framework in which artificial intelligence (AI) is tasked with autonomously discovering pulse sequences that realize spin-locked dynamics, even in the absence of analytical guidance.

### IV.1 Pulse sequence design via Proximal Policy Optimization

To overcome the limitations of analytical approaches in complex quantum systems, we turn to a data-driven strategy based on Reinforcement Learning (RL). RL provides a natural framework for control design problems in which an agent learns to make sequential decisions through trial and error. In our context, this means autonomously constructing pulse sequences that steer a many-body quantum system toward a desired dynamical behavior without requiring a closed-form solution for the control protocol.

We focus specifically on the Proximal Policy Optimization (PPO) algorithm, a modern and robust variant of policy-gradient methods[3,4]. In this setting, the agent observes a state — a mathematical representation of the current pulse sequence and its effect on the system — and takes an action, which corresponds to appending an additional pulse element. The environment updates accordingly, and the agent receives a reward that quantifies how effective the new sequence is in achieving a desired goal, such as approximating a target evolution. The agent's behavior is governed by a trainable policy, represented by a neural network known as the Actor, which proposes actions based on the current state. In parallel, a second network, the Critic, estimates the long-term value of that state to guide policy improvement. Over many iterations, the agent uses feedback from rewards and value estimates to refine both networks. To capture the sequential nature of pulse construction, we employ Long Short-Term Memory (LSTM) layers within both the Actor and Critic architectures. LSTM is a type of recurrent neural network architecture specifically designed to retain and update memory over long sequences, enabling the agent to make context-aware decisions based on the history of previously selected pulses.

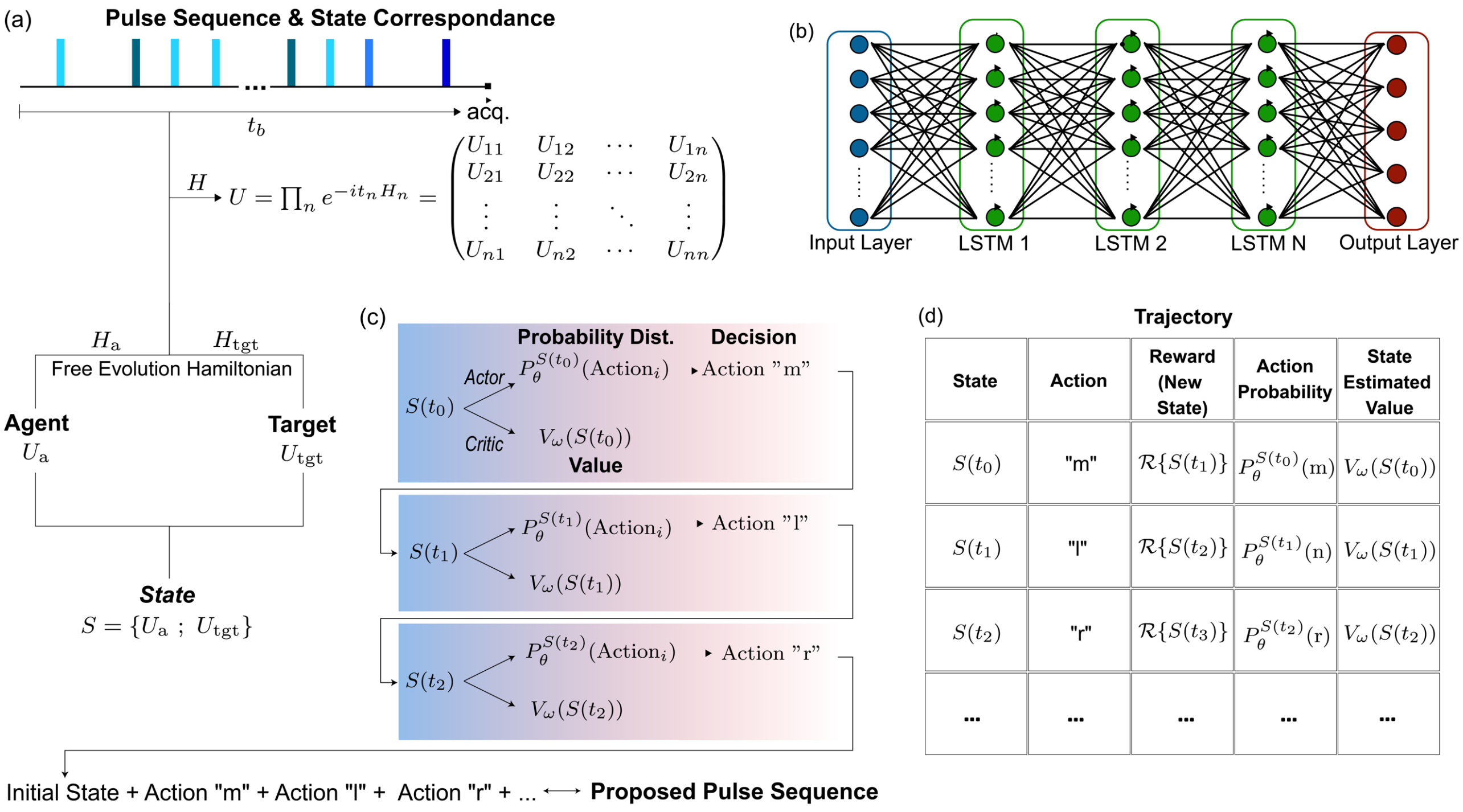

| State | Action | Reward (New State) | Action Probability | State Estimated Value |
|---|---|---|---|---|
| $S(t_0)$ | "m" | $\mathcal{R}\{S(t_1)\}$ | $P_\theta^{S(t_0)}(\text{m})$ | $V_\omega(S(t_0))$ |
| $S(t_1)$ | "l" | $\mathcal{R}\{S(t_2)\}$ | $P_\theta^{S(t_1)}(\text{n})$ | $V_\omega(S(t_1))$ |
| $S(t_2)$ | "r" | $\mathcal{R}\{S(t_3)\}$ | $P_\theta^{S(t_2)}(\text{r})$ | $V_\omega(S(t_2))$ |
| ... | ... | ... | ... | ... |

**Supplementary Figure 10 | PPO-based pulse sequence design.** (a) Mapping from pulse sequences to environment states, defined by the evolution operator $U_a$ for the physical system and a target operator $U_{tgt}$. (b) Neural network architecture for the PPO algorithm, comprising an Actor network (with trainable parameter set θ) that proposes actions and a Critic network (with parameters $\omega$) that evaluates state values. Both networks incorporate LSTM layers to retain contextual information across sequential decisions. The Actor's output layer parametrizes a state-dependent distribution probability $P_\theta^{S(t_i)}$ defined upon the action space. (c) Stepwise construction of a pulse sequence through state-dependent decisions. (d) Sample trajectory detailing state, action, reward, action probability, and value—used for training via stochastic gradient ascent.

The adaptation of PPO to pulse sequence design is summarized in Supplementary Fig. 10. As shown in Panel 10a, each sequence is mapped to a pair of evolution operators: $U_{\mathrm{a}}$, which captures the dynamics of the physical system under the proposed control, and $U_{\mathrm{tgt}}$, which encodes the target behavior. These two operators form the system's current state $S = \{U_{\mathrm{a}}, U_{\mathrm{tgt}}\}$. Based on this input, the Actor network proposes a next action — i.e., the next pulse to be added — by sampling from a probability distribution specific to that state (Supplementary Fig. 10b). The Critic network evaluates the expected long-term value of the state to guide optimization. This stepwise decision-making process is illustrated in Supplementary Fig. 10c. As the agent builds a complete sequence, it generates a trajectory — a record of states, actions, rewards, and probabilities — that is then used to update both networks via stochastic gradient ascent (Supplementary Fig. 10d).

Importantly, this approach enables exploration of pulse protocols in regimes where analytical tools are ineffective, whether due to complex many-body interactions, control constraints, or the lack of a known reference Hamiltonian. It also allows discovery of novel strategies tailored to arbitrary target dynamics, rather than relying on known sequences or symmetries.

### IV.2 AI-assisted pulse sequence design: Phase 1

The goal of Phase 1 is to autonomously generate compact pulse sequences that approximate a desired target evolution, such as effective decoupling in many-body quantum systems. This phase is particularly useful when no prior solutions are available in the literature, allowing the AI to bootstrap its learning without external guidance.

To enable tractable learning, we restrict the agent's action space to a minimal yet expressive set: Four "perfect" 90º pulses (with phases $\pm x$ and $\pm y$) followed by a free evolution time $\tau$, along with a fifth action representing free evolution of the same duration (i.e. no-pulse action). Each action thus appends a fixed-duration segment to the current sequence. While this limited set may appear restrictive, it facilitates faster convergence by narrowing the search space and encourages the discovery of WaHuHa-like building blocks, known to be effective for dipolar decoupling. To ensure sequences remain short and interpretable, each episode is capped at a maximum number of actions $N_{\mathrm{max}}$. This also justifies the separation between Phase 1 (which handles short sequences) and Phase 2 (which focuses on scalable repetition and symmetry exploitation).

Supplementary Figure 11 illustrates the core decision-making logic. At each step, the current pulse sequence defines the system state $S_i$. This state is processed by the Actor neural network, which outputs a probability distribution over available actions. One action is sampled stochastically and appended to the sequence. The updated sequence is then tentatively evaluated as a final proposal by computing a fidelity metric $\mathcal{F}(S_{i+1})$ between the actual and target evolution operators

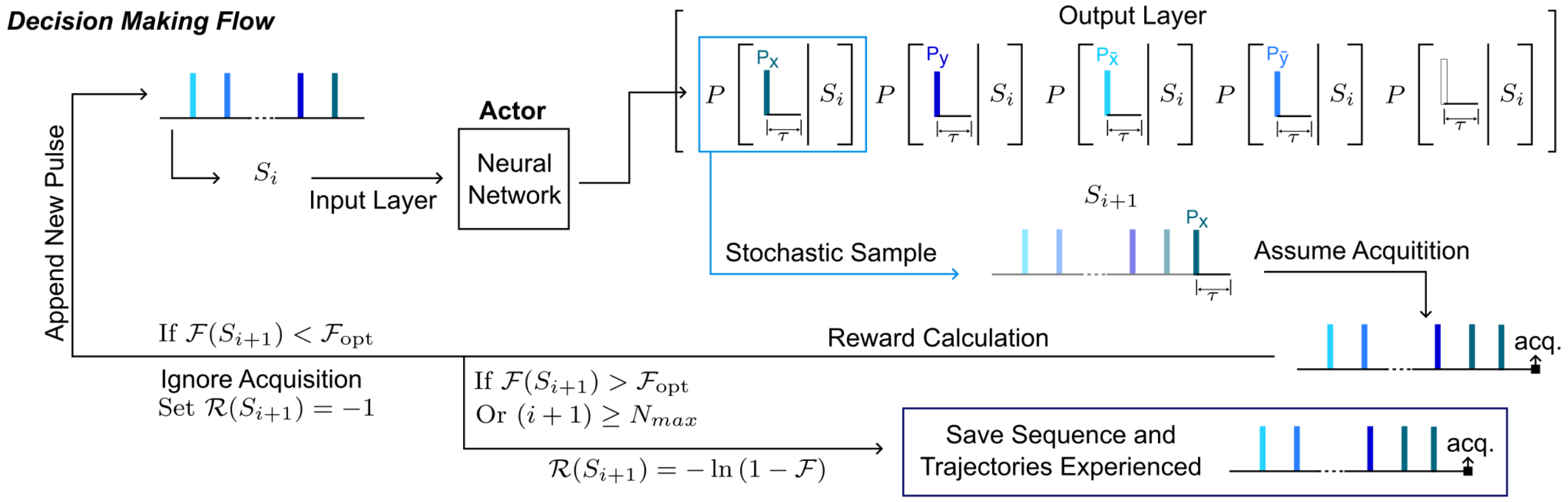

**Supplementary Figure 11 | Agent´s stochastic decision-making flow in Phase-1.** Starting from the top left and given a particular sequence (i.e., a particular state in the environment), the Actor neural network processes the state and parametrizes a distribution of probability. This distribution underlies the stochastic sampling of the action to be performed, which in this case is a "pulse + free evolution" to be appended to the current sequence. The following steps determine if the current sequence is efficient enough to be considered a final solution, or if the process should be continued, now starting from the current state.

$$\mathcal{F}(U_{\mathrm{a}}, U_{\mathrm{tgt}}) = \frac{1}{2^{\mathcal{N}}}\left(\mathrm{Tr}\left(\left|\sqrt{\widetilde{U}_{\mathrm{a}}\widetilde{U}_{\mathrm{tgt}}^{\dagger}}\right|\right)\right)^2, \tag{24}$$

where $\widetilde{U} = U/\sqrt{Tr\{UU^{\dagger}\}}$ denotes the normalized evolution operator for a system of $\mathcal{N}$ spin-1/2 particles. The fidelity serves as the basis for the reward function $\mathcal{R}(S_i)$, which guides learning:

$$\mathcal{R}(S_i) = \begin{cases} -\ln\big(1 - \mathcal{F}(S_i)\big), & \mathcal{F}(S_i) \geq \mathcal{F}_{\mathrm{opt}} \\ -1, & \mathcal{F}(S_i) < \mathcal{F}_{\mathrm{opt}} \end{cases} \tag{25}$$

Here, $\mathcal{F}_{\mathrm{opt}}$ is a predefined threshold used to terminate an episode early when a sufficiently good sequence is found. The logarithmic reward strongly favors high-fidelity outcomes while penalizing unnecessary sequence extensions.

Crucially, no individual pulse is inherently rewarded — only full sequences are evaluated — ensuring that the agent learns to value coherent combinations of operations over isolated actions. Although each intermediate step incurs a negative reward, successful final sequences provide a large enough payoff to reinforce the beneficial trajectories. This allows the neural network to refine its policy over time via stochastic gradient ascent, gradually increasing the probability of selecting actions that lead to high-fidelity control protocols.

### IV.3 Phase 2: Construction of long-lived control protocols from symmetry-driven building blocks

Once short, symmetry-informed pulse sequences have been identified in Phase 1, Phase 2 focuses on scaling these units into extended protocols capable of sustaining observable magnetization over much longer timescales. To this end, the action space is redefined to allow the AI agent to autonomously generate composite sequences by iteratively appending and transforming these basic building blocks.

The available actions fall into three categories (Supplementary Fig. 12a). *Type-0* actions allow the agent to append any one of the predefined blocks without regard to the current state. *Type-1* and *Type-2* actions apply transformations — such as sign reversals, phase shifts, or symmetrization rules — to the most recently added block(s), creating new candidates designed to suppress undesired dynamics through toggling-frame symmetry. These localized transformations enable efficient exploration of the combinatorial space while keeping the action set tractable. Although the approach could be extended to include deeper history (e.g., *Type-3* or *Type-4* actions), we restrict the transformations to one or two previous blocks to maintain efficient training.

The reward structure is also adapted to reflect the goals of Phase 2. Whereas in Phase 1 only the final state received a meaningful reward, in Phase 2 each newly proposed block is evaluated immediately based on how well it extends the protocol's coherence-preserving behavior. This is done using a fidelity-based metric, as before, but now with two additional hyperparameters: The target extension $t_{\mathrm{tgt}}$, which sets the intended protocol duration, and the projection time $t_{\mathrm{pj}}$, which estimates long-term performance by assuming the protocol is periodically repeated.

Supplementary Figure 12b summarizes the decision-making flow. Starting from a short initial protocol, the agent chooses its next action using the Actor neural network, which outputs a probability distribution over the available transformations. The selected action produces a new protocol, which is evaluated and either accepted for continuation or terminated if the total duration exceeds $t_{\mathrm{tgt}}$. In that case, the protocol is projected forward to $t_{\mathrm{pj}}$ by repeated application, and its total fidelity reward accumulated over multiple acquisition windows. This design favors protocols whose internal structure leads to stable, repeatable dynamics, allowing the AI to discover sequences that outperform traditional approaches over extended durations.

### IV.4 AI-guided protocol discovery and illustration: The aDSL-67 sequence

Efficient training of the two-phase AI framework depends critically on the choice of hyperparameters. Some are intrinsic to the PPO algorithm itself — such as neural network architecture, learning rate, or the number of epochs — while others are specific to the pulse-sequence design problem, including the structure of the reward functions and control over protocol duration. Optimizing across this multidimensional space is computationally intensive and generally problem-specific.

To address this, we adopted a parallelized approach: Instead of tuning a single AI instance, we trained multiple independent agents across a broad sweep of hyperparameters. Each training run becomes a guided stochastic search, and rather than preserving the final neural network weights, we retain only the best-performing pulse protocols discovered during training. In this way, the AI acts as a self-contained generator of novel solutions, without requiring further deployment post-training. Future work may revisit network fine-tuning or policy generalization across tasks.

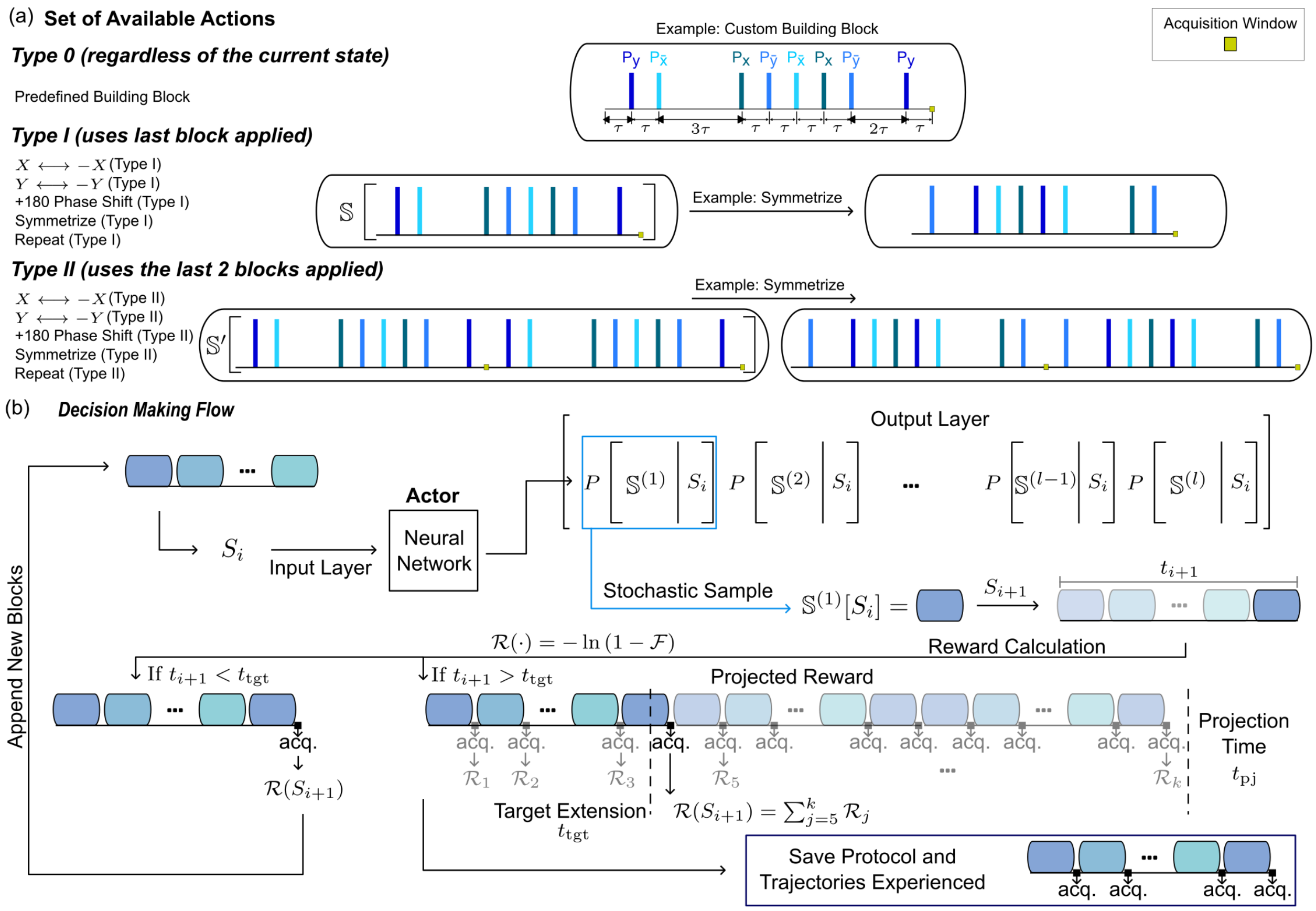

**Supplementary Figure 12 | Phase-2 action space and decision-making process.** (a) Classification of available actions used to build long-lived sequences from symmetry-informed building blocks. Type-0 actions append a predefined block regardless of the current state. Type-1 and Type-2 actions apply transformations (e.g., axis inversion, phase shift, or toggling-frame symmetrization) to the last one or two blocks, respectively, enabling efficient local exploration of the protocol space. Each block ends in an acquisition window (acq.). (b) Schematic of the agent's decision-making flow. Given a current state , the Actor neural network samples a transformation to apply from a state-dependent probability distribution. If the resulting protocol remains shorter than a predefined target extension $t_{\text{tgt}}$, the fidelity-based reward is computed and a new action is sampled. Once the target duration is exceeded, a projected reward is calculated by periodically repeating the protocol up to a projection time $t_{\text{pj}}$, and summing the fidelity over all acquisition windows. This favors protocols that sustain observable magnetization across long timescales.

This strategy proved effective. In Phase 1, multiple high-fidelity blocks were discovered, many resembling known sequences such as WaHuHa. Notably, the AI independently converged on cyclic sequences — even though cyclicity was not imposed by design — enabling post-hoc analytical treatment via AHT. These blocks served as the basis for constructing longer sequences in Phase 2, where the agent combined and transformed them to generate extended protocols exhibiting robust spin locking.

One illustrative result is the aDSL-67 sequence, consisting of 67 decision steps beginning from a single Phase-1-designed building block. The agent combined symmetry operations and repetitions to build a protocol containing 536 pulses and 67 acquisition windows, with a total duration $t_c^{\text{AI}} = 804\tau$ (Supplementary Fig. 13). The intuition behind the selection of such a long pulse sequence is that even small detunings from resonance should be efficient in locking spins. As in the case of DSL-4, however, the standard treatment of using the Magnus series to capture the dynamics of the one-cycle evolution operator $U(t_c) = \prod_{k=1}^{n} P_k e^{-i\tau\mathcal{H}} = \prod_{k=1}^{n} e^{-i\tau\mathcal{H}'_k}$ proves inadequate. Specifically, in the absence of dipolar couplings an expansion of the aDSL-67 protocol to second order yields

$$\mathcal{H}_{\text{M}}^{(0)} = \omega\left(\frac{1}{3}I_z + \frac{7}{201}I_y\right) \;;$$

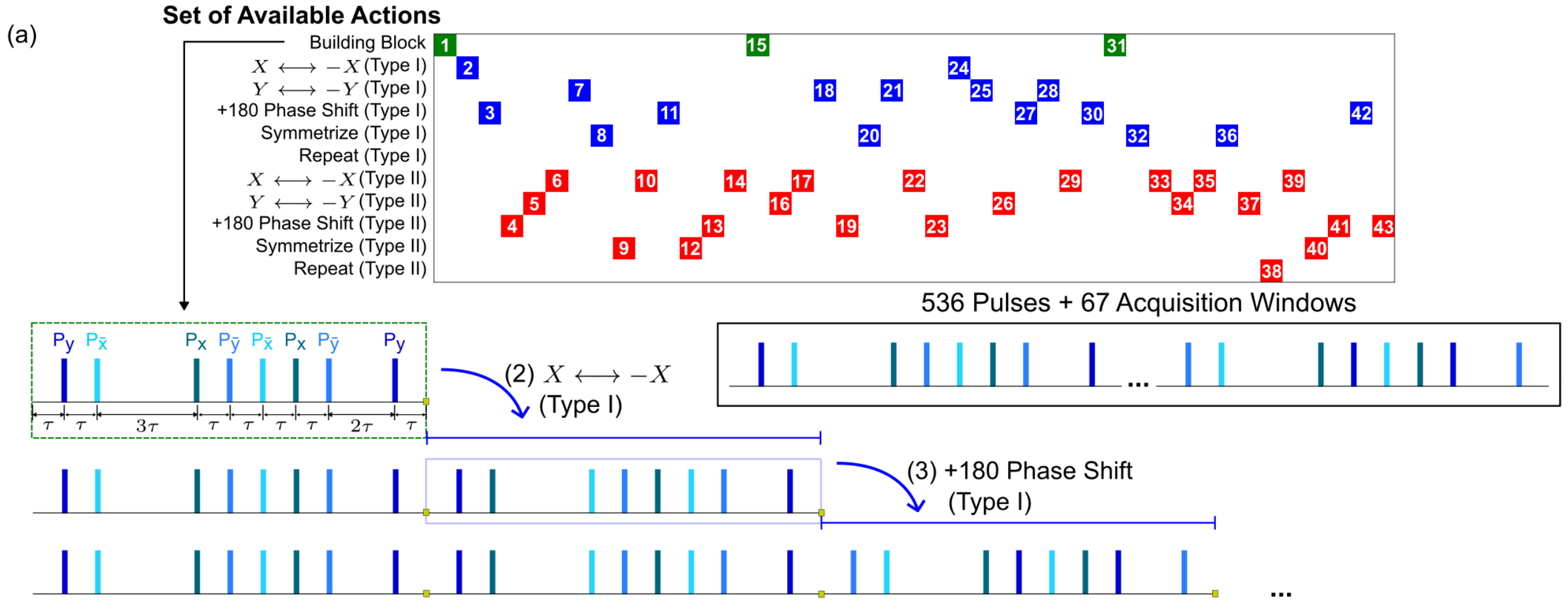

**Supplementary Figure 13 | Emergence of protected dynamics from an AI-designed pulse protocol.** Definition of the aDSL-67 pulse sequence generated during a Phase-2 episode. The protocol begins from a Phase-1-designed building block and evolves through 67 symmetry-based transformations (Type-1 and Type-2), leading to 536 pulses (tall blueish rectangles) and 67 acquisition windows (yellow squares). The first few steps are shown explicitly.

$$\mathcal{H}_{\mathrm{M}}^{(1)} = \tau\omega^2\left(\frac{15}{268}I_z + \frac{1}{67}I_y - \frac{515}{268}I_x\right) \;;$$
$$\mathcal{H}_{\mathrm{M}}^{(2)} = \tau^2\omega^3\left(-\frac{41405}{4824}I_z - \frac{67489}{2412}I_y - \frac{801}{536}I_x\right) \;; \tag{26}$$

which we compare to the exact dynamics in Supplementary Fig. 14a for the simplified case of non-interacting spins whose initial magnetization points along $x$. The failure of average Hamiltonian theory becomes evident even for modest detunings within ±1 kHz: The central spectral peak at zero frequency, a hallmark of offset-induced locking, is absent in both the first- and second-order Magnus approximations (Supplementary Figs. 14b and 14c). At larger offsets, the disagreement worsens, again exposing the inability of perturbative AHT to capture dynamic spin locking.

The offset-adapted framework of Section I, however, allows us to better characterize the aDSL-67 dynamics. Specifically, Supplementary Fig. 15a shows that both the amplitude and tilt of $\delta$ vary sharply with detuning, giving rise to offset windows where the locking torque is maximized or suppressed. Importantly, these variations occur on the scale of only a few kilohertz, highlighting the fine spectral structure that can emerge for long-cycle sequences. The influence of detuning extends beyond the effective field to impact the magnitude of the offset-dressed dipolar Hamiltonian, $\|\widetilde{\mathcal{D}}_{\rm M}\|$, which inherits a pronounced dependence on $\nu$. In particular, simulating the dynamics of a 10-spin dipolar-coupled cluster — where, as before, we quantify the locking efficiency $\mathcal{L}_S(\nu)$ through the amplitude of the resonance peak at the locking frequency — reveals structured dips delimiting regimes where spin locking is preserved or suppressed (lower plot in Fig. 4b of the main text).

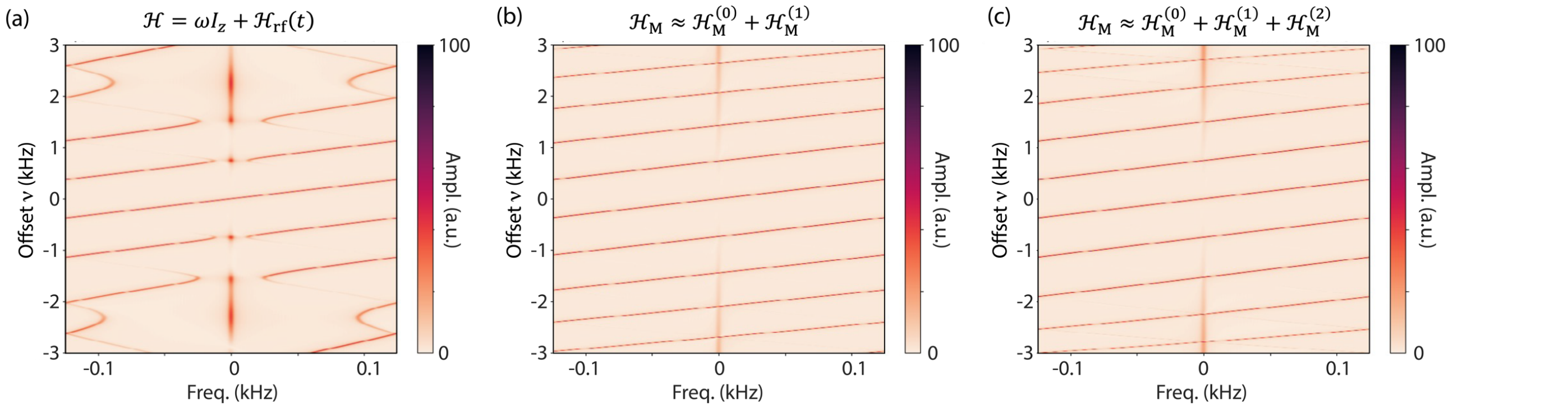

**Supplementary Figure 14 | Breakdown of average Hamiltonian theory under detuning for the aDSL-67 sequence.** Fourier transforms of the magnetization response are shown as a function of offset frequency $\nu$ for evolution under the aDSL-67 pulse protocol, starting from $\rho(0) = I_x$ and for stroboscopic acquisition limited to once per cycle. (a) Exact frequency-domain response in the absence of dipolar interactions under the time-dependent Hamiltonian $\mathcal{H} = \omega I_z + \mathcal{H}_{\mathrm{rf}}(t)$. (b) Same as in (a) but using an effective Hamiltonian including terms in the Magnus expansion up to first-order, and (c) to second-order, both computed analytically without including dipolar interactions. While higher-order corrections improve the agreement slightly, neither approximation captures the true spectral features beyond modest detunings ($|\nu| \gtrsim 700$ Hz). The discrepancy highlights the breakdown of perturbative methods for long-cycle sequences, even in the absence of interactions.

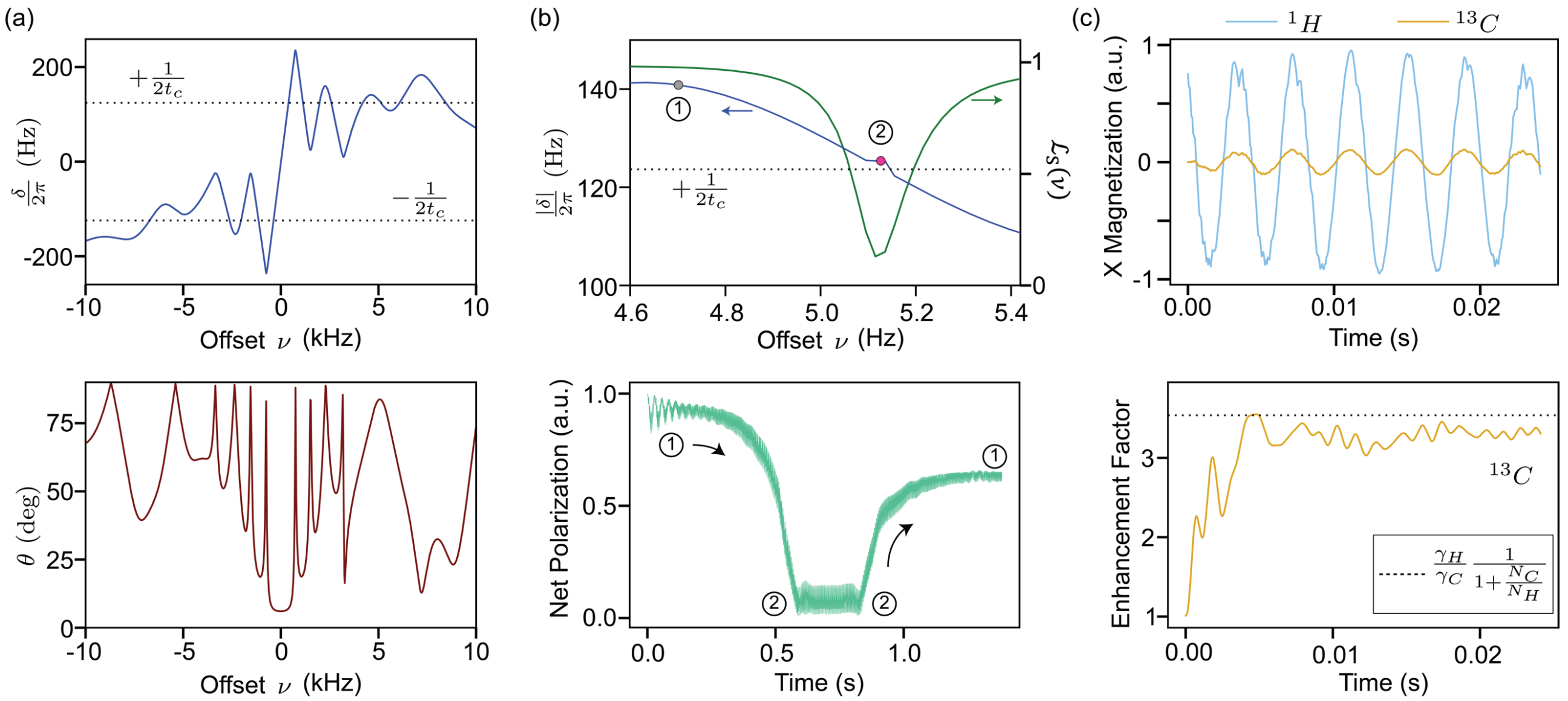

**Supplementary Figure 15 | Adiabatic demagnetization and polarization transfer under the aDSL-67 protocol.** (a) Offset dependence of the effective locking field $\delta(\nu)$, shown via its amplitude (top) and tilt angle relative to $z$ (bottom) for $\tau = 5$ μs. (b) Top: Zoomed view of the locking field amplitude (blue trace) and locking efficiency $\mathcal{L}_S(\nu)$ (green trace); the gray and red dots respectively indicate the starting and end frequencies of a demagnetization and remagnetization cycle. Bottom: Calculated net polarization of a 7-proton spin cluster as one slowly varies the offset frequency; the polarization recovers only partially due to the finite frequency sweep rate. (c) Top: aDSL-mediated polarization transfer as calculated from a spin cluster comprising eight $^1H$ and one $^{13}C$ nuclei for rf pulse durations of 1 and 2 μs, respectively, away from the Hartmann-Hahn condition. To mitigate symmetry-driven oscillations, we combine the responses from ten spin clusters with identical spin composition but different spatial arrangement; for these simulations, $\nu = -733$ Hz, and the dipolar couplings in the spin cluster range from 0.5 to 3 kHz. Bottom: Calculated $^{13}C$ polarization enhancement as a function of time; after a short transient, the $^{13}C$ polarization stabilizes into a value close to the theoretically predicted maximum (dotted horizontal line).

Supplementary Fig. 15b brings these ingredients together: By numerically sweeping the detuning across a prominent locking dip, we simulate a full demagnetization–remagnetization cycle for a 7-proton spin cluster under aDSL-67 control. The top panel shows the locking field amplitude (blue curve) and the locking efficiency $\mathcal{L}_S(\nu)$ (green curve), which drops sharply near $\nu \approx 5.1$ kHz. When the offset is swept slowly from an initial value (gray dot) through the dip and back (red dot), the system follows a largely reversible trajectory, with the net magnetization decreasing and then recovering over the course of the sweep (lower panel). To maintain the time required for these simulations within practical bounds, we limited the spin number to only 7, and implemented a varying sweep rate, designed to decrease when the derivative magnitude of $r(\nu) = \|\widetilde{\mathcal{D}}_M\|/|\delta|$ is greatest; inevitably, however, the finite sweep rate induces partial nonadiabaticity near the locking dip and leads to a partial polarization loss. This limitation, however, is not fundamental: Slower sweeps would reduce such losses and enable more complete recovery, suggesting that high-fidelity entropy control via aDSL-based control is within reach.

In addition to demagnetization, aDSL sequences can be used to mediate polarization transfer across spins with distinct gyromagnetic ratios far from the Hartmann-Hahn matching condition. To illustrate this point, Supplementary Fig. 15c presents numerical results from a mixed spin system consisting of eight protons and one carbon-13 nuclei. Here, protons and carbons are driven with $\pi/2$-pulse durations of 1 and 2 μs, respectively, away from the Hartmann–Hahn condition. Despite this mismatch, aDSL-67 introduces offset-dressed interaction pathways that allow for polarization buildup in the carbon spins. As shown in the lower panel, the $^{13}C$ polarization rapidly approaches a steady-state value comparable to the theoretical maximum expected from regular cross-polarization (dotted line), achieving efficient transfer without the need for resonant driving.